\documentclass[12pt,preprint]{aastex}

\usepackage{epsfig}

\shorttitle{Planetary Nebulae in GLIMPSE~3D}
\shortauthors{Zhang et al.}

\begin{document}

\title{Planetary Nebulae Detected in the {\it Spitzer Space Telescope} GLIMPSE~3D
Legacy Survey}

\author{Yong Zhang, Chih-Hao Hsia, Sun Kwok}
\affil{Department of Physics, The University of Hong Kong, Hong Kong
\email{zhangy96@hku.hk,xiazh@hku.hk,sunkwok@hku.hk}}

\begin{abstract}

We used the data from the {\it Spitzer}  Legacy Infrared Mid-Plane Survey Extraordinaire (GLIMPSE) to investigate the mid-infrared (MIR)  properties of planetary nebulae (PNs) and PN candidates.
In previous studies of GLIMPSE~I $\&$ II data, we have shown that these MIR data are very useful in distinguishing PNs from other emission-line objects. In the present paper, we focus on the PNs in the field of the GLIMPSE~3D survey, which has a more extensive latitude coverage. We found a total of 90 Macquarie-AAO-Strasbourg (MASH) and MASH~II PNs and 101 known PNs to have visible MIR counterparts in the GLIMPSE~3D survey area.  The images and photometry of these PNs are presented.
Combining the derived IRAC photometry at 3.6, 4.5, 5.8, 8.0\,$\mu$m with the existing photometric measurements from other infrared catalogs, we are able to construct spectral energy distributions (SEDs) of these PNs.  Among the most notable objects in this survey is the PN M1-41, whose GLIMPSE~3D image reveals a large bipolar structure of more than 3 arcmin in extent.

\end{abstract}

\section{INTRODUCTION}

Planetary nebulae (PNs)  are important tools for the study of stellar nucleosynthesis
and galactic chemical evolution. However, due to interstellar extinction, current optical census of Galactic PNs is highly incomplete, and some compact \ion{H}{2} regions might have been mis-classified as PNs. The problem is particularly severe in the Galactic plane where the interstellar extinction is significant, and consequently the detections of PNs at optical bands are severely hampered. Unlike optical observations, infrared  (IR) observations are only marginally affected by interstellar extinction and observations of PNs in the IR is highly desirable.

PNs are bright in the IR due to their circumstellar dust component.  The early prediction that the remnant of the circumstellar dust envelope from the asymptotic giant branch (AGB) progenitors should still be observable in the PN phase \citep{kwok82} has been confirmed by the 
 {\it Infrared Astronomical Satellite} ({\it IRAS}) all-sky survey  (Pottasch et al. 1984; Zhang $\&$ Kwok 1991).  The peak of the infrared emission lies between 20-30 $\mu$m, corresponding to a color temperature of $\sim$100-150 K.  Quite often a significant fraction of the energy output of PNs is emitted in the IR and mid-IR brightness has become a defining characteristics of PNs.  

Although the structure of the ionized component of PNs is well determined by optical observations, early IR observations have too low spatial resolution to resolve the structures of the dust component of PNs.  The recent {\it Spitzer Space Telescope} and the Galactic Legacy Infrared Mid-Plane Survey Extraordinaire (GLIMPSE; Benjamin et al. 2003, Churchwell et al. 2009)  mapped the inner Galaxy at mid-infrared (MIR) 3.6 , 4.5, 5.8, and 8.0\,$\mu$m using the Infrared Array Camera \citep[IRAC;][]{fazio04}.  With high resolution and sensitivity, the GLIMPSE data provide new insights into the nature of the nebular components contributing to the infrared emission of PNs.  The GLIMPSE~I $\&$ II images cover a total area of about 274\,deg$^2$ of the Galactic plane and these data have been used for the studies of PNs  \citep[e.g.][]{cohen07b,cohen11,phillips08a,phillips08b,phillips09,ramos08}.
Our research group has carried out systematic searches of PNs in the fields of GLIMPSE~I $\&$ II \citep{kwok08,zhang09}.
Our previous results show that the IR appearances of PNs might differ from their optical counterparts, and these IR images can help to distinguish between PNe and \ion{H}{2} regions.

In the present work  we present a study of PNs in the area of the GLIMPSE~3D survey. The GLIMPSE~3D program extends the GLIMPSE~I $\&$ II coverage to higher latitude, where the IR background is not so bright as in  the low-latitude regions.  
Although \citet{quino11} have investigated the IR properties of 24 PNs in the GLIMPSE~3D survey area, our sample is much larger (with a total of 191 PNs).  
This paper should be considered as a complement of our previous work \citep{kwok08,zhang09}.

\section{OBSERVATIONS AND DATA REDUCTION}

This study is primarily based on IRAC images, centered at approximately 3.6, 4.5, 5.8
and 8.0\,$\mu$m, from the GLIMPSE~3D dataset \citep{mea07}. All the four bands
were simultaneously observed with a pixel resolution of $\sim1.2\arcsec$.
The GLIMPSE~3D survey extends the GLIMPSE~I $\&$ II latitude coverage 
to  $\left|b\right|<3^\circ$  at nine selected strips above and below the Galactic plane
(centered at $l=10, 18.5, 25, 30, 330, 335, 341.5, 345,$ and 350$^\circ$),
and to $\left|b\right|<4.2^\circ$  in the center of the Galaxy ($\left|l\right|<2^\circ$).
The total area is about 120\,deg$^2$. Figure~\ref{cover} shows the area coverage of the 
GLIMPSE~3D survey.  

The GLIMPSE~3D data products are very similar to those of  GLIMPSE~I $\&$ II. The basic
data calibration was performed by the Spitzer Science Center (SSC). The basic calibrated
data were further processed using the GLIMPSE pipeline to correct for further instrumental artifacts, cross-identify and determine the flux densities and positions of point sources, and mosaic the images. Further details of the GLIMPSE archive and data processing can be found in \citet{ben03}. The  GLIMPSE~3D data products include a highly reliable Point Source Catalog (GLM3DC),  a more complete Point Source Archive (GLM3DA), and mosaic images covering the survey areas.

The sample of PNs was taken in part from the MASH (Macquarie/AAO/Strasbourg H$\alpha$ Planetary Galactic Catalogue) and MASH~II catalogues \citep{parker06,miszalski08}, which contain over 1000 new faint PNs and PN candidates (hereafter, simply referred to as MASH PNs) discovered in the  AAO/UKST H$\alpha$ survey of the southern Galactic plane. The sample contains some of the most obscured PNs in the Galaxy. The survey area of GLIMPSE~3D survey is completely covered by the AAO/UKST H$\alpha$ survey. Furthermore, we also searched for  all the previously known PNs catalogued by \citet{koh01} within the GLIMPSE~3D survey area.  These PNs are designated as ``previously known PNs'' in the following discussion.

We visually examined the IRAC images of all these PNs lying within the 
GLIMPSE~3D area. If a PN is visible in at least one of the bands, we 
measured its integrated fluxes using the same method as in our previous
papers \citep{kwok08,zhang09}. Two apertures were employed to measure
the on-source and background fluxes ($F_{on}$ and $F_{off}$). In all the
four IRAC bands, the apertures were put in the identical positions.
We then obtained the sum of all the fluxes of sources in the point-source 
archive within each aperture ($F_{on,p}$ and $F_{off,p}$).
Finally, the PN fluxes were determined from $(F_{on}-F_{on,p})-(F_{off}-F_{off,p})$.
For each PN,  we measured the fluxes using several apertures with different
sizes, and the adopted values are the average of all the measurements.
From these repeated measurements, we estimate the typical error of 
the photometry to be $5\%$, and up to $30\%$ for the sources with low surface brightness.
Because the radii of these PNs are typically larger than $8''$,
we made extended source aperture correction using the correction factors
suggested by Jarrett\footnote{http://spider.ipac.caltech.edu/staff/jarrett/irac/calibration/ext\_apercorr.html}.

We also used the 24\,$\mu$m data taken by the Multiband Imaging Photometer 
for {\it Spitzer} \citep[MIPS;][]{rieke04}.  The data were retrieved from
the {\it Spitzer} Legacy program MIPS Inner Galactic Plane Survey (MIPSGAL).
The MIPSGAL survey only has a small overlap in Galactic coverage with the 
GLIMPSE~3D survey. 
We determined the 24\,$\mu$m fluxes of PNs using the same method as described above. To better explore nebular IR emission, we also made use of data from other IR archives, including the Two Micron All Sky Survey (2MASS), Deep Near Infrared Survey of the Southern Sky (DENIS), {\it Midcourse Space Experiment (MSX)}, the {\it IRAS} Point Source Catalogue (PSC), and specially the recently released AKARI Point Source Catalogues.   In the AKARI All-Sky Survey \citep{mur07}, the mid- and far-infrared images were obtained using the Infrared Camera (IRC) and Far-Infrared Surveyor (FIS) with spatial resolutions from $4''$--$61''$.
 This study is also complemented by the  1.4\,GHz
radio fluxes taken from the NRAO VLA Sky Survey \citep[NVSS;][]{condon98}.

\section{RESULTS}

We find that there are 228 MASH PNs in the GLIMPSE~3D area, among which 90 have clearly visible IRAC counterparts.  The IRAC color composite images of these 90 PNs are shown in Figure~\ref{image}, where
 we can easily see that the colors of these PNs are redder than those of the field stars.   
The notes on individual MASH PNs are given in Appendix~A.  The majority of the GLIMPSE~3D PNs have a size of $<30$\,arcsec.   An inspection of Figure~\ref{image} clearly shows that the surface brightness of PNs is lower for larger PNs. For the MASH PNs that are very extended in the visible, their IR counterparts are difficult to find as the IR surface brightness are likely to fall below the IRAC detection limit.

\begin{figure*}
\epsfig{file=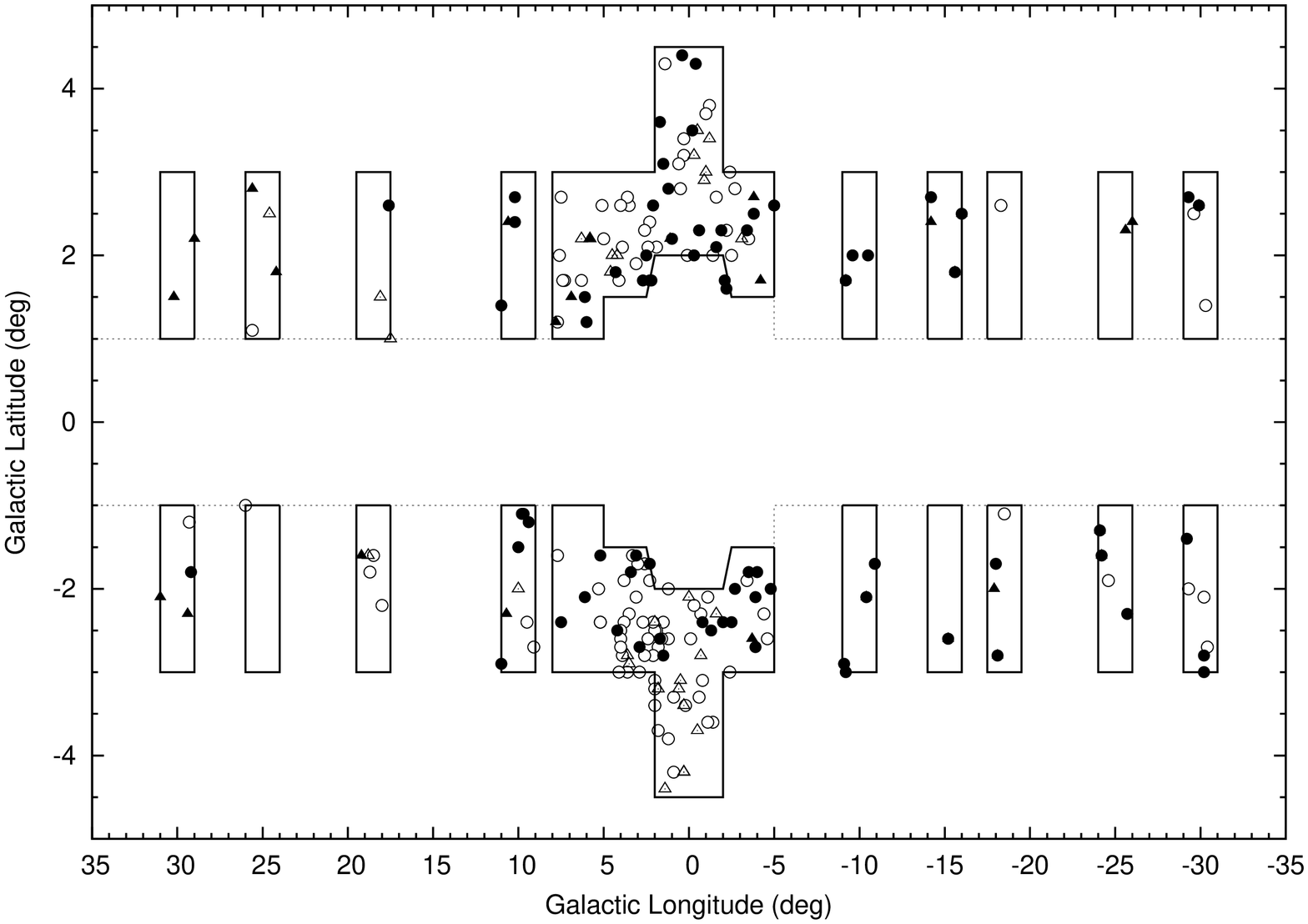, height=15cm, angle=-90}
\vspace{0.5in}
\caption{
The sky coverage of GLIMPSE~3D survey (zones enclosed by solid lines). The dotted lines delineate the survey areas of GLIMPSE I $\&$ II.  The locations of MASH and MASH~II PNs lying within the GLIMPSE~3D survey area are shown by circles and triangles, respectively.  The filled and open symbols respectively represent the PNs with and without IRAC counterparts.}
\label{cover}
\end{figure*}
\begin{figure*}
\begin{center}
\begin{tabular}{ccccc}
\resizebox{30mm}{!}{\includegraphics{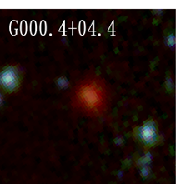}} &
\resizebox{30mm}{!}{\includegraphics{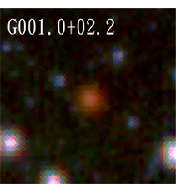}} &
\resizebox{30mm}{!}{\includegraphics{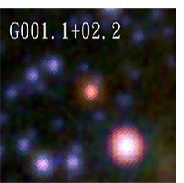}} &
\resizebox{30mm}{!}{\includegraphics{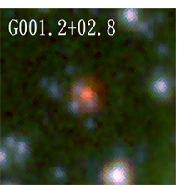}} &
\resizebox{30mm}{!}{\includegraphics{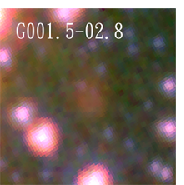}} \\
\resizebox{30mm}{!}{\includegraphics{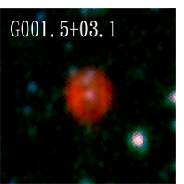}} &
\resizebox{30mm}{!}{\includegraphics{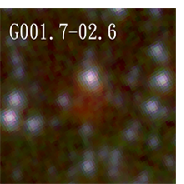}} &
\resizebox{30mm}{!}{\includegraphics{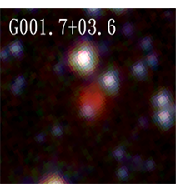}} &
\resizebox{30mm}{!}{\includegraphics{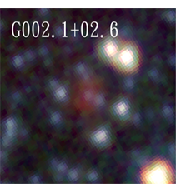}} &
\resizebox{30mm}{!}{\includegraphics{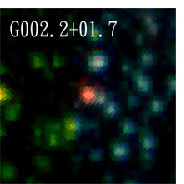}} \\
\resizebox{30mm}{!}{\includegraphics{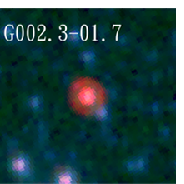}} &
\resizebox{30mm}{!}{\includegraphics{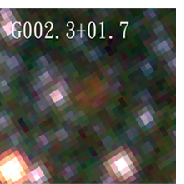}} &
\resizebox{30mm}{!}{\includegraphics{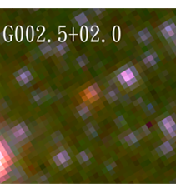}} &
\resizebox{30mm}{!}{\includegraphics{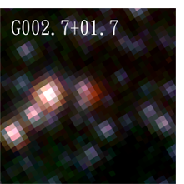}} &
\resizebox{30mm}{!}{\includegraphics{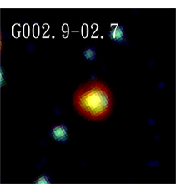}} \\
\resizebox{30mm}{!}{\includegraphics{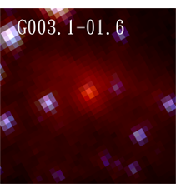}} &
\resizebox{30mm}{!}{\includegraphics{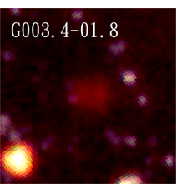}} &
\resizebox{30mm}{!}{\includegraphics{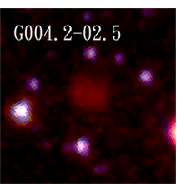}} &
\resizebox{30mm}{!}{\includegraphics{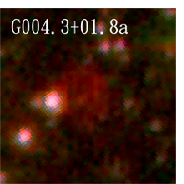}} &
\resizebox{30mm}{!}{\includegraphics{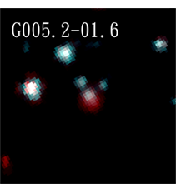}} \\
\end{tabular}
\end{center}
\caption{Composite-color images of 90 MASH PNs detected in the GLIMPSE~3D survey (abridged version).  The images were made from the three IRAC bands: 3.6\,$\mu$m (shown as blue), 5.8\,$\mu$m (green), and 8.0\,$\mu$m (red). Each panel covers an area of $40''\times40''$.
North is up, and east is to the left.  \label{image}}
\end{figure*}

Figure~\ref{cover} shows that the spatial density of GLIMPSE~3D PNs is larger near the Galactic center. The detection rate 
($R_{IR}$; the number of PNs having visible IR counterparts over the total number) of the GLIMPSE~3D PNs is $42\%$, larger than
that of GLIMPSE~I PNs which are located in the inner Galactic plane \citep{kwok08}. There are two main factors affecting the detection rate. Because the spatial distribution of the galactic dust is more diffuse than the ionized gas (which is more concentrated around hot stars), the effect of background emission on the detection of PNs at IR wavelengths is larger than that at optical.  As a result, the bright IR background emission in the inner Galactic plane tends to decrease the $R_{IR}$ value for the PNs close to the Galactic plane. On the other hand, unlike the H$\alpha$ emission, the IR emission of PNs are hardly obscured by the interstellar dust. This tends to increase the  $R_{IR}$ value for the PNs near the Galactic plane where the interstellar extinction is larger than the outer regions. Indeed, as shown  in Figure~\ref{cover}, the $R_{IR}$ value  in the regions of $\left|b\right|>3^\circ$ is obviously smaller than that in the inner
regions. 

\begin{figure*}
\begin{center}
\begin{tabular}{ccccc}
\resizebox{30mm}{!}{\includegraphics{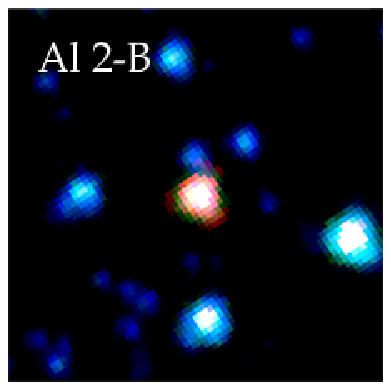}} &
\resizebox{30mm}{!}{\includegraphics{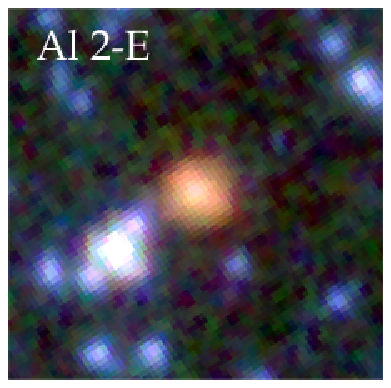}} &
\resizebox{30mm}{!}{\includegraphics{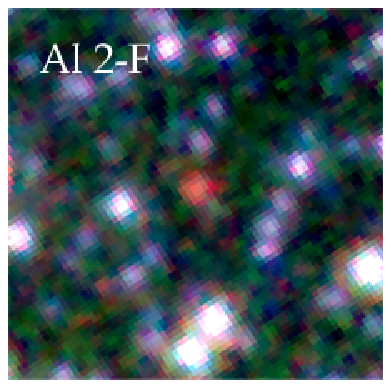}} &
\resizebox{30mm}{!}{\includegraphics{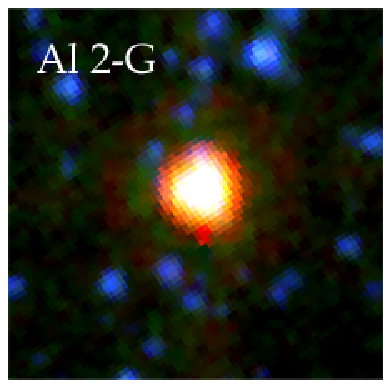}} &
\resizebox{30mm}{!}{\includegraphics{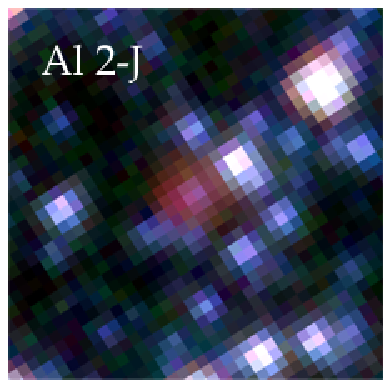}} \\
\resizebox{30mm}{!}{\includegraphics{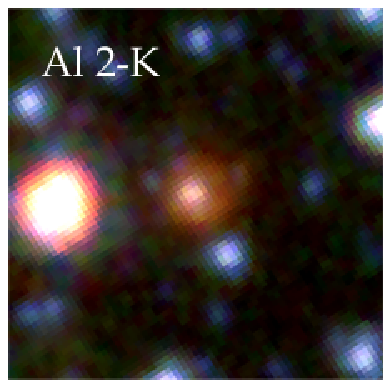}} &
\resizebox{30mm}{!}{\includegraphics{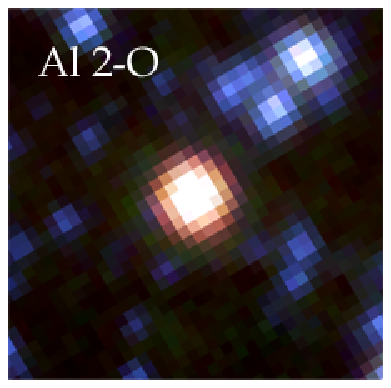}} &
\resizebox{30mm}{!}{\includegraphics{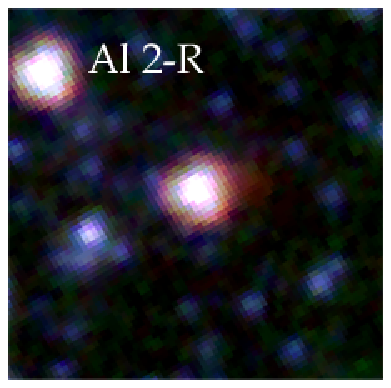}} &
\resizebox{30mm}{!}{\includegraphics{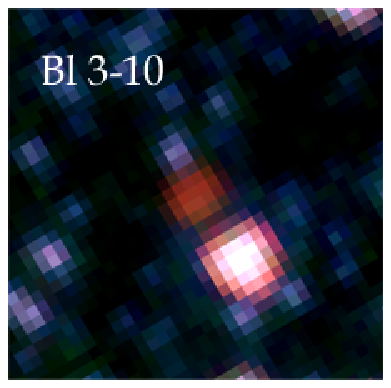}} &
\resizebox{30mm}{!}{\includegraphics{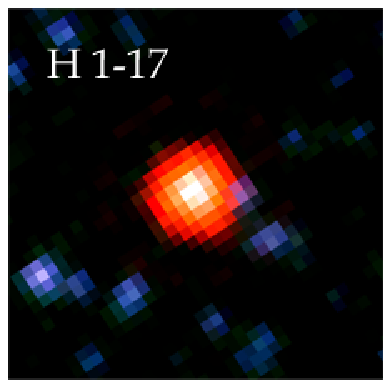}} \\
\resizebox{30mm}{!}{\includegraphics{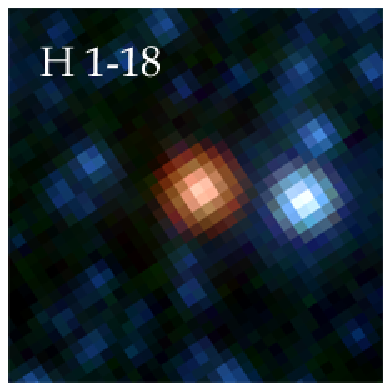}} &
\resizebox{30mm}{!}{\includegraphics{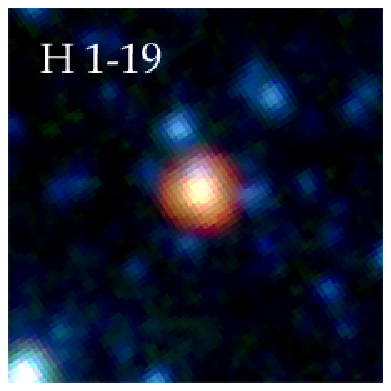}} &
\resizebox{30mm}{!}{\includegraphics{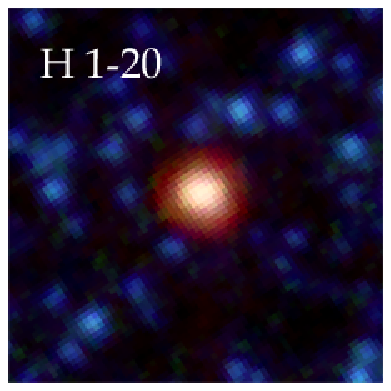}} &
\resizebox{30mm}{!}{\includegraphics{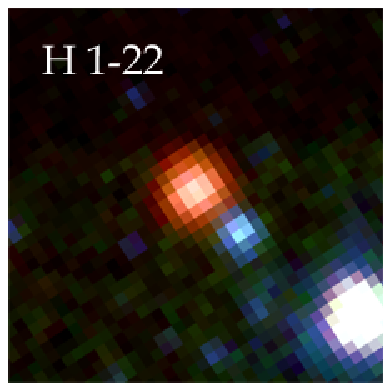}} &
\resizebox{30mm}{!}{\includegraphics{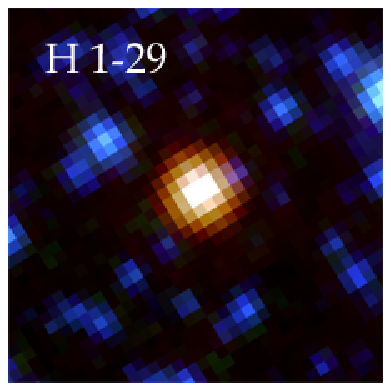}} \\
\resizebox{30mm}{!}{\includegraphics{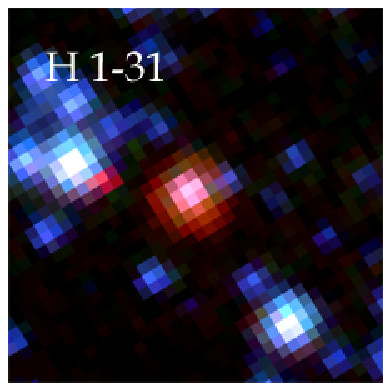}} &
\resizebox{30mm}{!}{\includegraphics{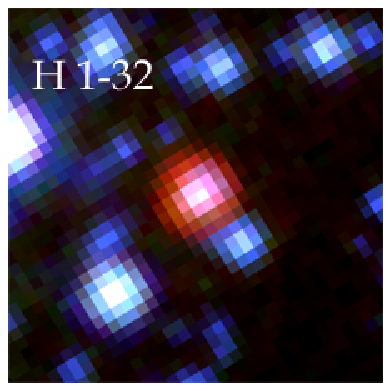}} &
\resizebox{30mm}{!}{\includegraphics{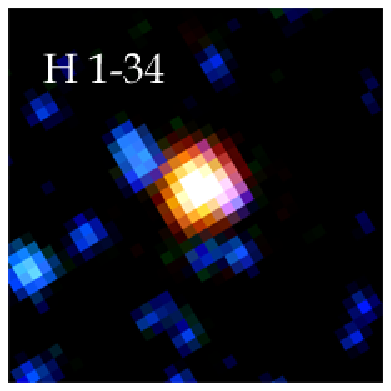}} &
\resizebox{30mm}{!}{\includegraphics{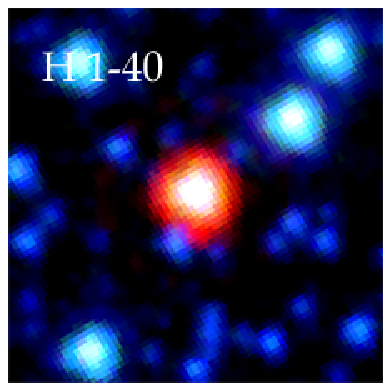}} &
\resizebox{30mm}{!}{\includegraphics{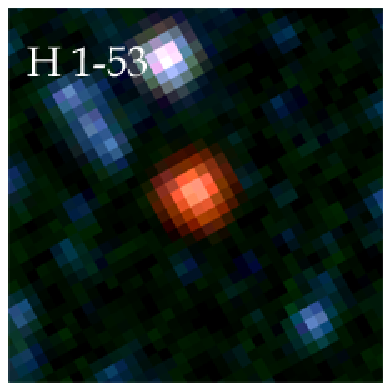}} \\
\end{tabular}\end{center}
\caption{Composite-color images of 80 previously known PNs detected in the GLIMPSE~3D survey (abridged version).  The images were made from the three IRAC bands: 3.6\,$\mu$m (shown as blue)
, 5.8\,$\mu$m (green), and 8.0\,$\mu$m (red). Except those marked, each panel covers an area of $40''\times40''$.
North is up, and east is to the left.  \label{knownpn}}
\end{figure*}

Among 107 previously known PNs, 101 have obvious IRAC counterparts. The high
detection rate of previously known PNs is because they are in general brighter than the MASH PNs.
We have carefully examined the GLIMPSE images of these 101 PNs, some show similar appearances as the optical images, and some only reveal the brighter central regions of the objects.  Since they are often bright, some are saturated at the 8.0 $\mu$m band.
The IRAC color composite images of 80 previously known PNs are presented in Figure~\ref{knownpn}, where we have excluded the saturated sources and those with
non-detection in some bands. \citet{quino11} and \citet{phillips08a} presented the IRAC images of several known PNs, which are also excluded from Figure~\ref{knownpn}.
We have also attempted to search for fainter outer structures but this effort is often hindered by strong infrared background. A  previously known PN (M~1-41) is discussed in detail in Section~3.3.

 

\subsection{SPECTRAL ENERGY DISTRIBUTIONS}

The Spitzer IR fluxes of the GLIMPSE~3D PNs (including MASH and previously known PNs) are given in Table~\ref{flux}.  About 37\% of the GLIMPSE~3D PNs lie within the MIPSGAL/24\,$\mu$m survey area. We find that the PNs are generally bright in the  24\,$\mu$m band. 
We also search for the counterparts of these PNs in the AKARI All-Sky Survey.  Table~\ref{akari} gives the IRC and FIS fluxes of the GLIMPSE~3D PNs. Table~\ref{other} tabulates the magnitudes and  fluxes from  the DENIS, 2MASS, MSX, IRAS, and NVSS point source catalogues. In far-IR wavelengths, the spatial resolutions  are low ($>30''$), and the field stars cannot be resolved from the PNs.
However, these PNs are much brighter than the surrounding stars at longer 
wavelengths.  We infer that the contamination from the field stars to the far-IR fluxes should be negligible.

\begin{figure*}
\begin{center}
\begin{tabular}{cc}
\resizebox{90mm}{!}{\includegraphics{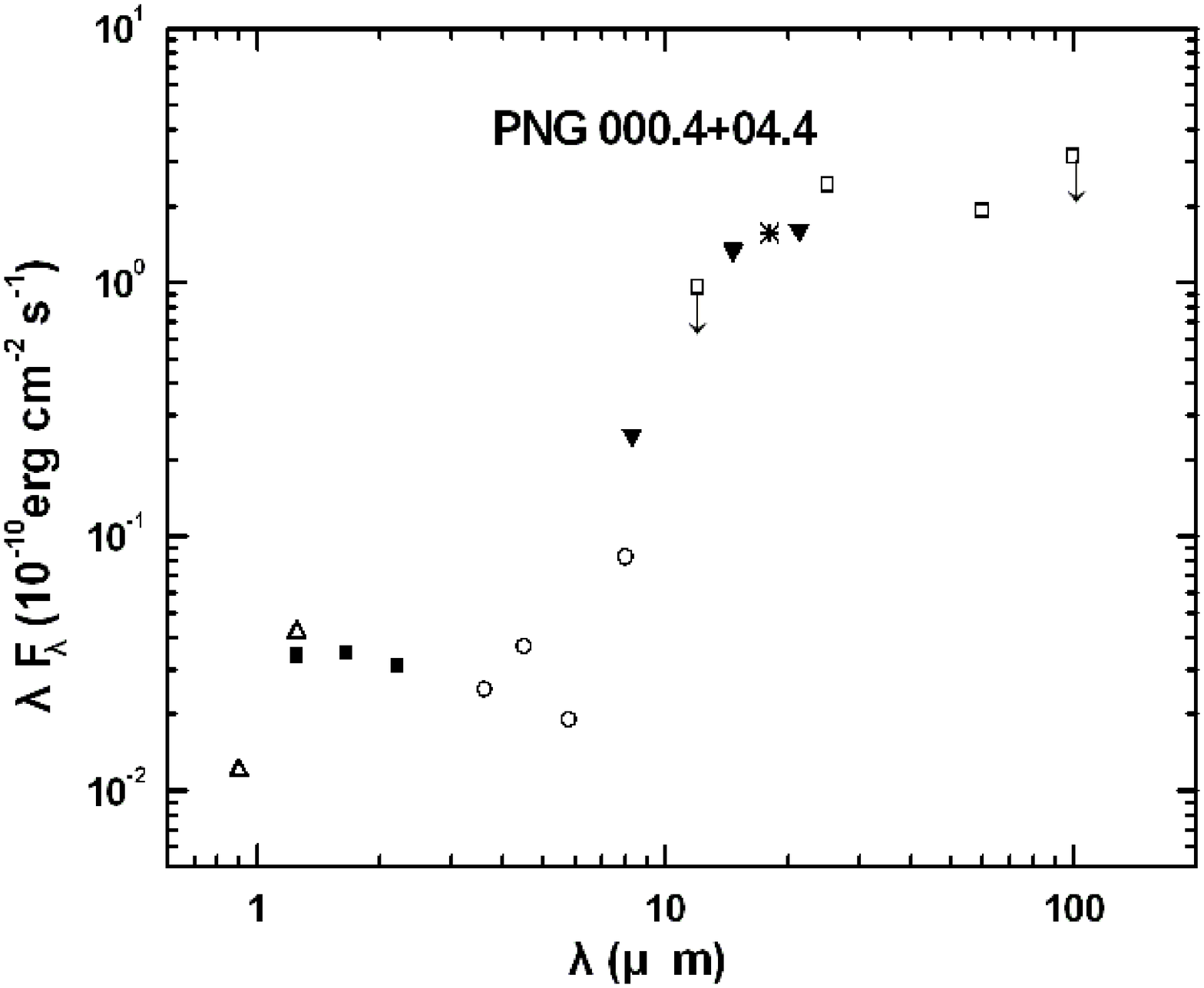}} & \resizebox{90mm}{!}{\includegraphics{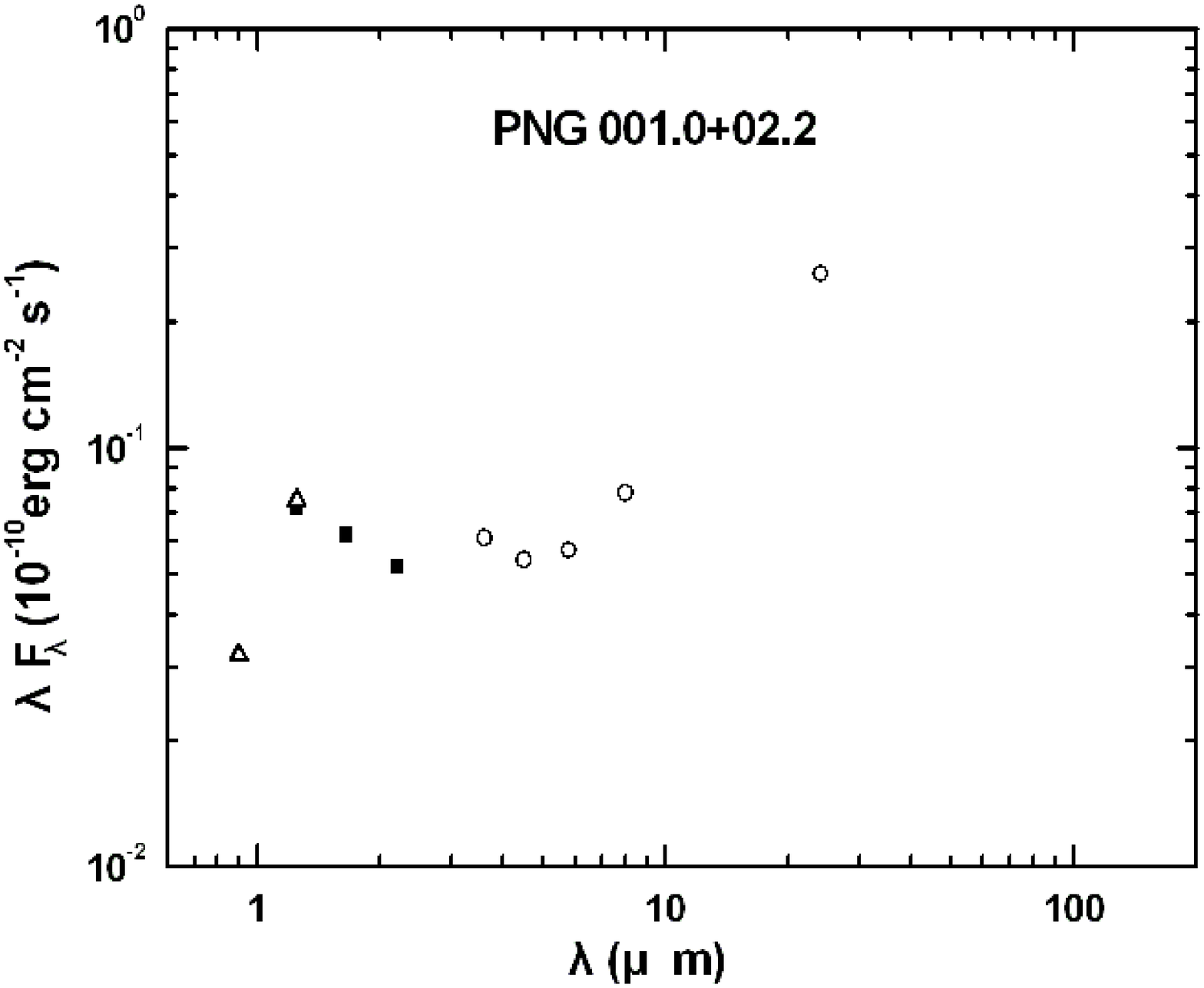}} \\
\end{tabular}
\end{center}
\caption{The SEDs of 83 PNs in the GLIMPSE~3D sample (abridged version). The open triangles,
filled squares, open circles, open squares, filled triangles, and asterisks
are from the  DENIS, 2MASS, GLIMPSE/MIPSGAL, {\it IRAS},  {\it MSX},
and AKARI survey, respectively. The light asterisks represent the
uncertain AKARI detections. The ISO and Spitzer spectra  of some PNs are also overlaid.
The dotted lines represent the blackbody curve with temperatures indicated.
\label{sed}
}
\end{figure*}

The SEDs of the GLIMPSE~3D PNs are constructed using the IR data from various databases (Figure~\ref{sed}). Since the IR data cover a wide wavelength range, we are able to derive the color temperatures of the dust component (e.g. see the case for NGC\,6302). For most of the PNs, there is a rise in flux toward short wavelengths. This is due to the contributions from the photospheric and nebular bound-free emission 
(see Zhang \& Kwok 1991 for a detailed discussion of SEDs of PNs)
and in some cases from a warm dust component.   
For a few previously known PNs, the ISO and Spitzer/IRS spectra are available and these are plotted in Figure~\ref{sed}.  In general, the spectra are in good agreement with the IR photometry data.  The SEDs at long wavelengths  ($>10$\,$\mu$m) can be reasonably approximated  by  blackbodies of temperatures $\sim100$\,K, although the fluxes are likely to be contaminated by the emission from aromatic infrared bands (AIB) and emission lines.

\begin{figure*}
\epsfig{file=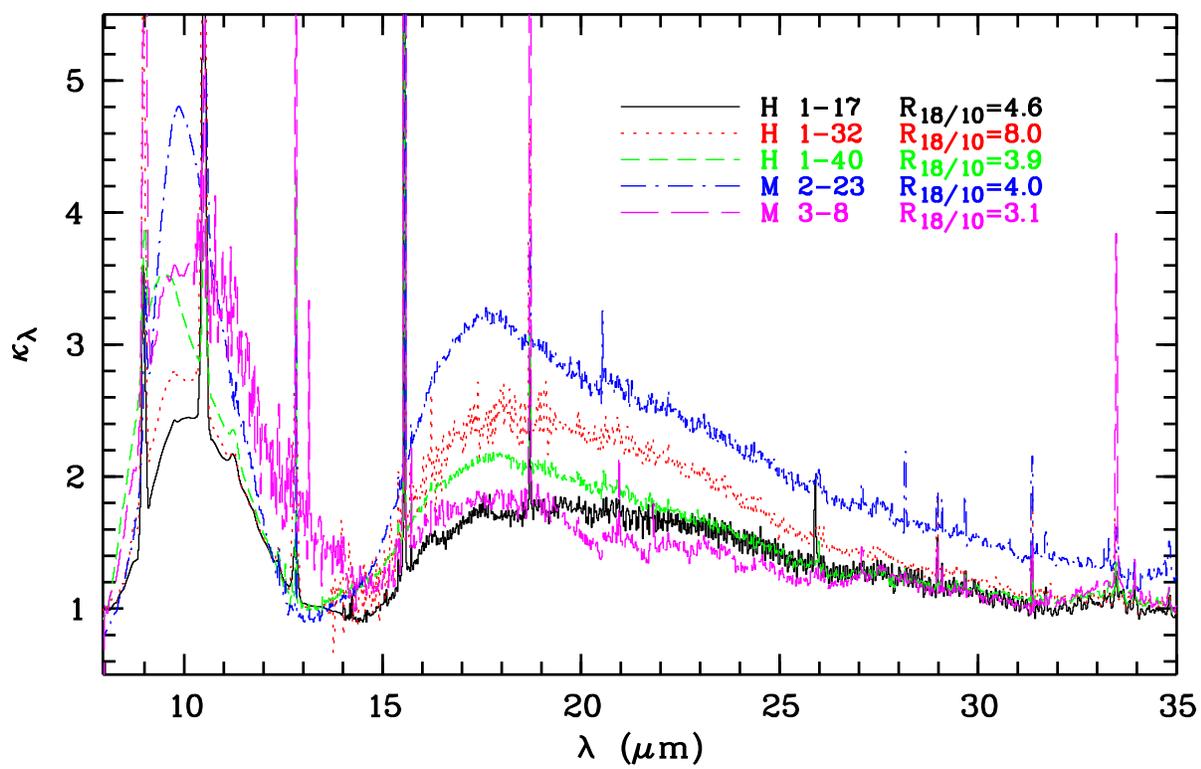, height=10cm}
\caption{Emissivities of the PNs exhibiting strong 10 and 18\,$\mu$m
features.
\protect\label{kap}}
\end{figure*}

Some PNs (e.g. H~1-17, H~1-32, H~1-40, M~2-23, and M~3-8) display prominent emission features from amorphous silicate grains at 10 and 18\,$\mu$m. All of them are non-Type~I PNs and show an O-rich or mixed chemistry.  The shapes and relative strengths of the two silicate features reflect the properties of the grains \citep[e.g.][]{sim91,don94}.
For our sample, we find the strength ratios of the 18 and 10\,$\mu$m ($R_{18/10}$) lie
within the range of 3--8. The variation of the $R_{18/10}$ value may be due to different volume fractions of graphite or porosity of the grains in different sources \citep{vai11}. The derived $R_{18/10}$ values are about one order of magnitude higher than those suggested by models of \citet{vai11}. However, we should mention that the uncertainties of subtracting the continuum may introduce large errors in measuring the strengths of the broad 18\,$\mu$m feature.  In Figure~\ref{kap}, we compare the scaled emissivities, $\kappa_\lambda=F_\lambda/B_\lambda(T)$, where $B_\lambda(T)$ is
the blackbody function with a temperature of $T$. Although we do not find obvious difference in the feature profiles between different sources, the peak position of the 10\,$\mu$m feature shifts
from source to source, probably indicating to the variety of chemical composition of the grains \citep[e.g.][]{min07}. The peak wavelength of the 10\,$\mu$m feature ranges from 9.5 to 10.3\,$\mu$m, and is shortest in H~1-40. We also find that H~1-40 has a mixed chemical composition and display strongest AIBs among the five PNs, supporting the conclusion of \citet{vai11} that the 10\,$\mu$m feature
shifts shortwards with graphite inclusions.

It is clear from the SEDs that in many PNs there are more than one dust component.  While the cool dust components (dominant at wavelengths $>$10 $\mu$m) are well defined, there are excesses between 5 and 10 $\mu$m which is most likely to be due to a warm dust component (see examples of H 1-16, H 1-18, H 1-19, H 1-34 in Fig.~\ref{sed}).  High spatial resolution observations are needed to identify the origin of this warm dust component.

\subsection{Infrared COLORS}

\begin{figure*}
\epsfig{file=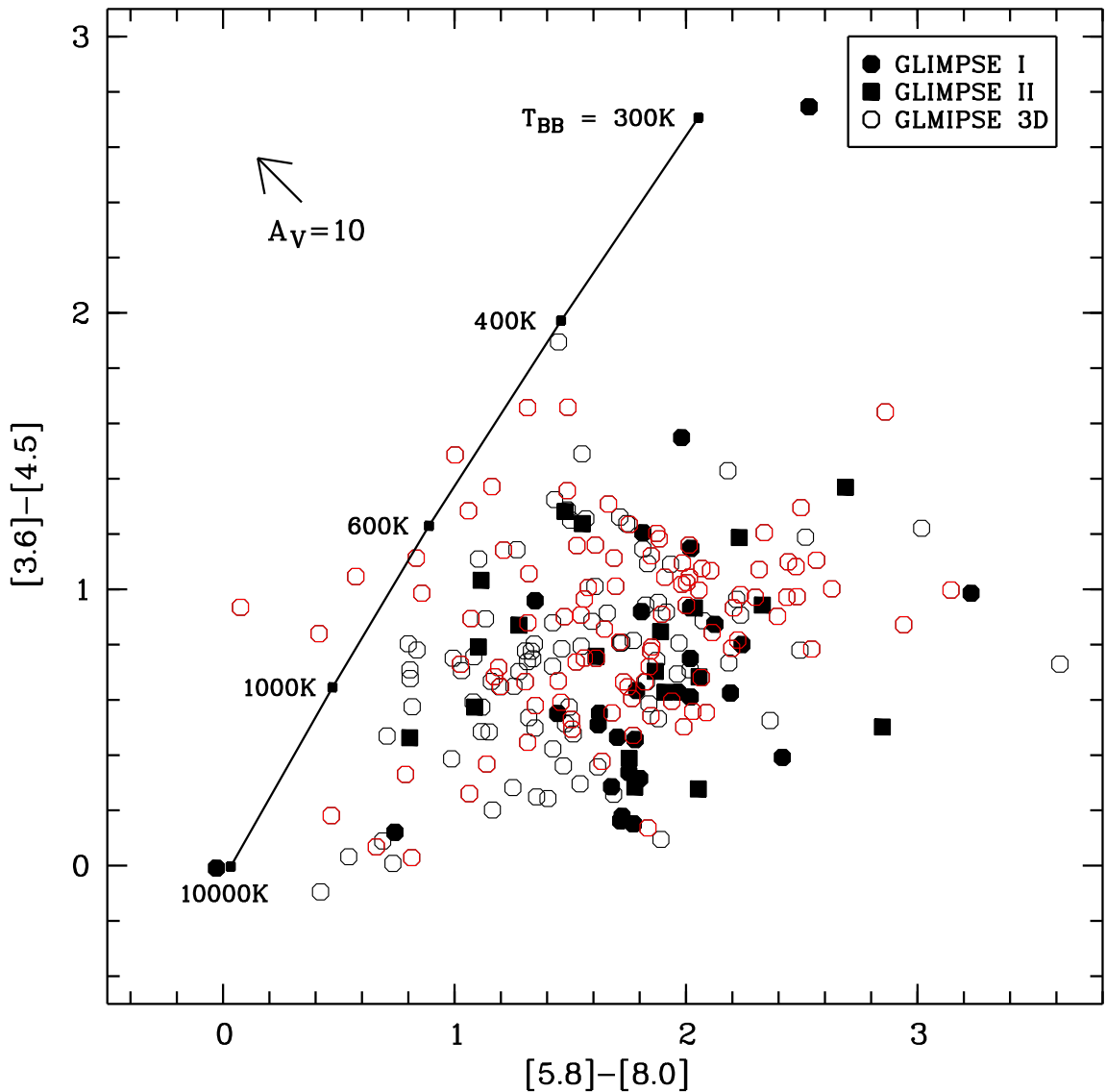, height=15cm}
\caption{IRAC color-color plot ($[3.6]-[4.5]$ vs. $[5.8]-[8.0]$)
for 231 GLIMPSE PNs with good fluxes at all 4 IRAC bands. The red and black symbols denote the previously known and MASH PNs, respectively.  The solid line is a track of blackbodies at temperatures $T_{\rm BB}$. The arrow in the upper left corner denotes a reddening vector of ${\rm A_v}=10$.
\protect\label{color}}
\end{figure*}

In Figure~\ref{color}, we compare the distributions of GLIMPSE PNs in the $[3.6]-[4.5]$ versus  $[5.8]-[8.0]$ color-color diagram. We do not find obvious differences between the colors of GLIMPSE~3D and GLIMPSE~I/II PNs.
Since a variety of emission mechanisms contribute toward the fluxes of the IRAC bands, we do not expect the color distribution to obey the blackbody law.  These contributing factors include the cool dust component (probably the most dominant), the warm dust component, nebular gas emission, AIB emissions, and even photospheric emissions.
Generally speaking, the sample PNs have colors to the right and below the blackbody line.  For objects with IRAC fluxes dominated by cool dust emission, this trend can be explained by emissivity dependence on wavelength, which makes the longer wavelength bands fainter.
Based on a study of 24 PNs, \citet{quino11} found that the previously known PNs tend to have a
larger $[5.8]-[8.0]$ color, and suggested that this is an evolutionary consequence.
However, our large sample study does not suggest such a trend (see Figure~\ref{color}).
For the PNs both in the samples of \citet{quino11} and this paper, we derive an average $[5.8]-[8.0]$ 
color of 1.95, in agreement with the value of GLIMPSE~I PNs obtained by \citet{cohen11}, but
lower than that by \citet{quino11} ($\sim2.38$). This might be in part due to different
aperture correction factors used for extended source calibration. On the other hand,
for the sources with large sizes, we only measure the bright central regions, 
probably resulting in different colors with those derived by \citet{quino11}.

\begin{figure*}
\epsfig{file=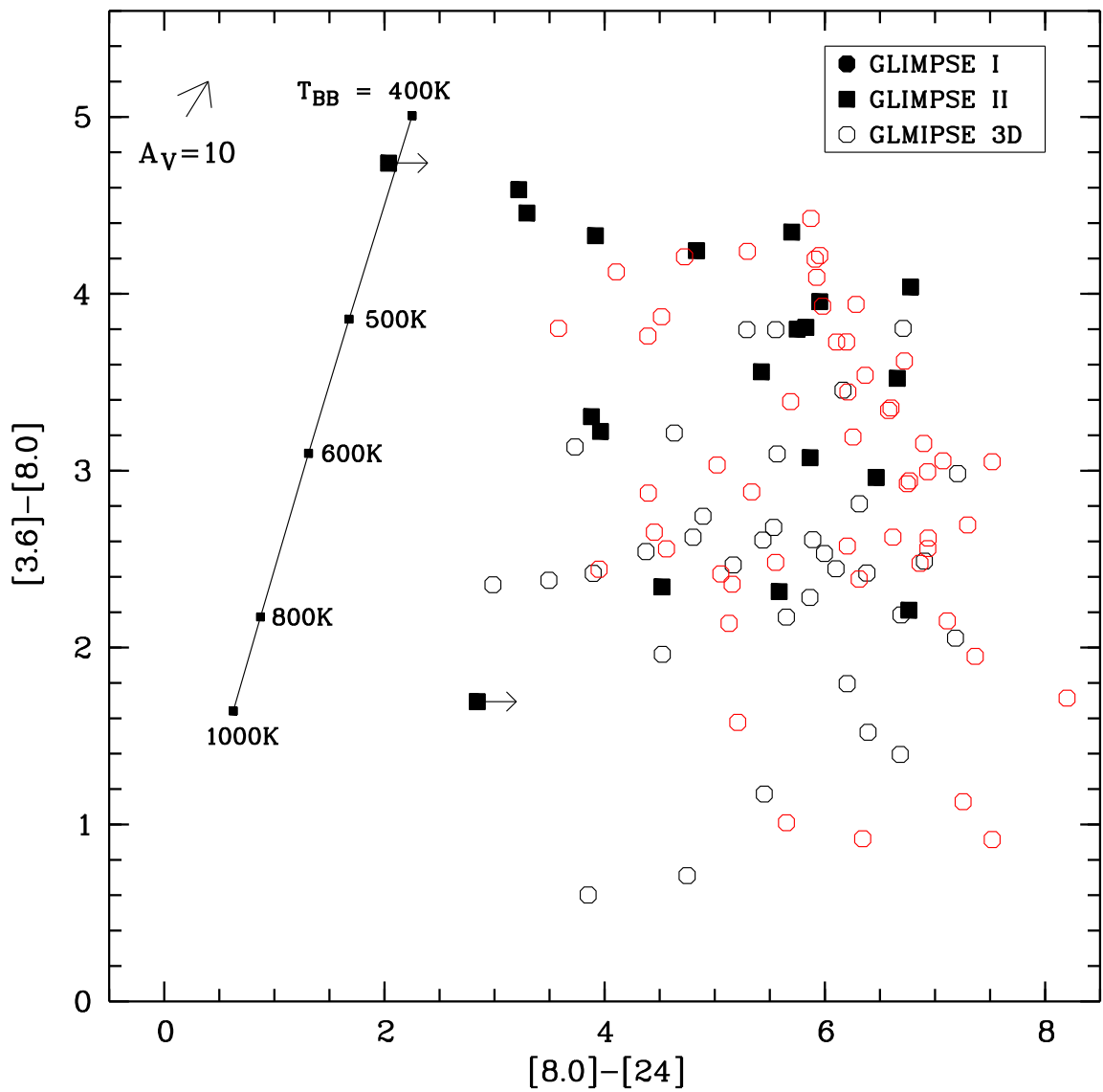, height=15cm}
\caption{The $[3.6]-[8.0]$ versus $[8.0]-[24]$ color-color
diagram for 106 GLIMPSE PNs with good fluxes at 3.6, 8.0, and 24 $\mu$m. Symbols are otherwise same as in Figure~\ref{color}.
\protect\label{color2}}
\end{figure*}

\begin{figure*}
\centering
\epsfig{file=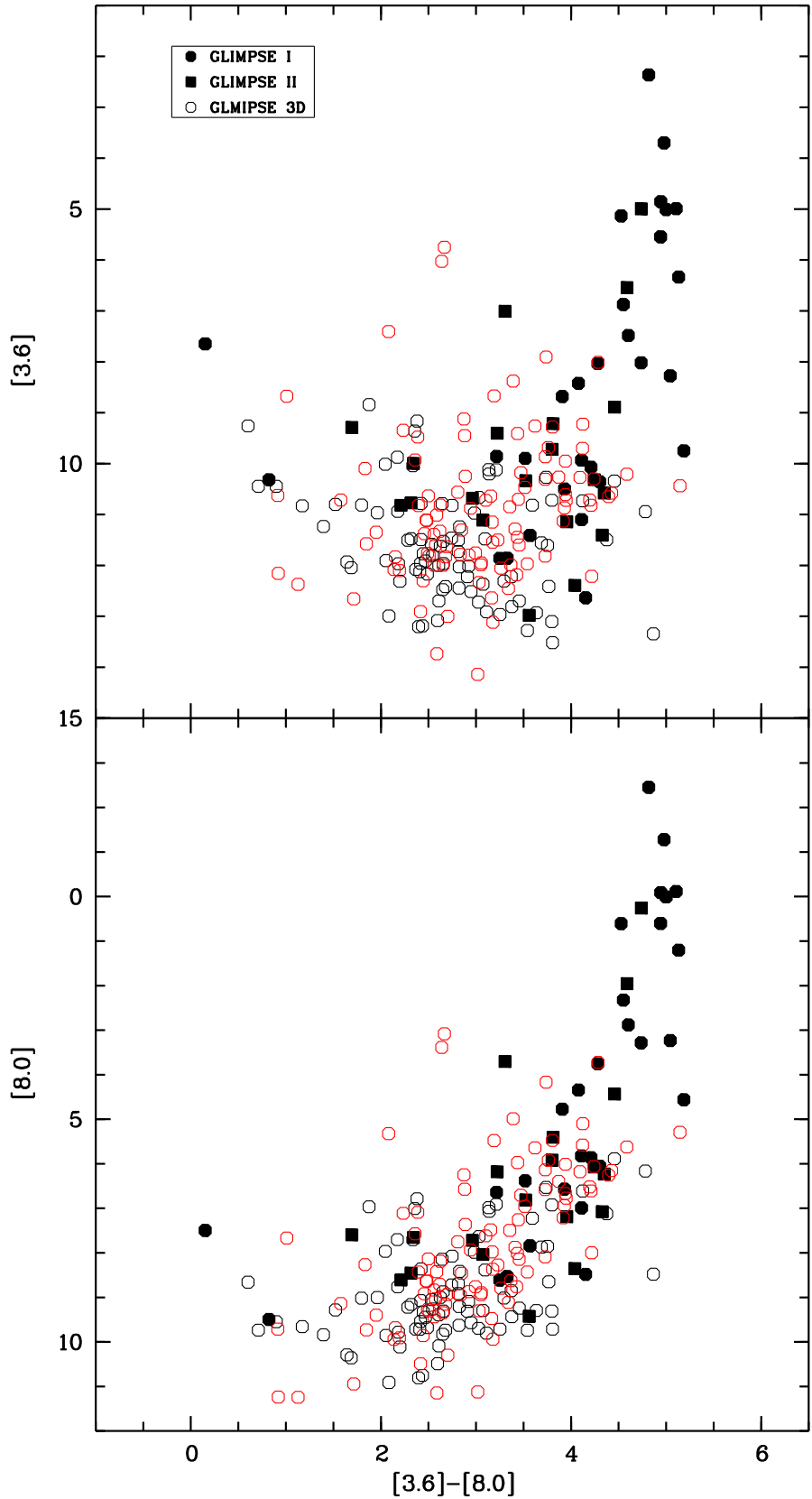, height=22cm}
\caption{ The [3.6] versus $[3.6]-[8.0]$ (upper panel)
and  [8.0] versus $[3.6]-[8.0]$  (lower panel) color-magnitude diagrams for 231 GLIMPSE PNs.
Symbols are otherwise same as in Figure~\ref{color}.
\protect\label{colmag}}
\end{figure*}

In Figure~\ref{color2}, we plot the $[3.6]-[8.0]$ versus  $[8.0]-[24]$ color-color diagram. 
The $[3.6]-[8.0]$ colors of GLIMPSE~3D PNs seem to be systematically smaller than
 those of GLIMPSE~II PNs.   This trend can also be seen in Figure~\ref{colmag} which gives the $[3.6]$ and $[8.0]$  versus $[3.6]-[8.0]$ color-magnitude diagrams.
In Figure~\ref{colmag} we can also find that the IR emissions of GLIMPSE~3D PNs are generally fainter than those of GLIMPSE~I/II PNs. This is an indication that in the GLIMPSE~I and II survey areas the detection of the PNs with intrinsically fainter IR emission is severely hampered by the bright background emission in the inner galactic plane.
Based on GLIMPSE~I data, \citet{cohen11} found that the MASH~II PNs have a smaller $[4.5]-[5.8]$ color than the MASH~I PNs,
and suggested that the MASH~II PNs are more compact and younger.
According to our measurements of GLIMPSE~3D, the average $[4.5]-[5.8]$ values are 0.44 and 0.42 for MASH~I and II PNs, respectively.
Our results suggest that in the higher-latitude regions where there dust exctinction is lower,
the bias in the discovery of MASH~I and II PNs is smaller than that in the inner galactic plane.

\citet{cohen11} found a trend that the $[4.5]-[5.8]$ and $[5.8]-[8.0]$ color indices 
of GLIMPSE~I PNs increases and decreases with PN age (proportional to intrinsic sizes of PNs), 
respectively, suggesting that the relatively strengths of AIBs change as PN age. 
In the GLIMPSE~3D survey area, most of the MASH PNs are compact and it is hard
to estimate their sizes. To examine the  relation between infrared colors and
PN age, we divide our sample into two groups: I) small sources with clear boundary; II)
extended sources or those with diffuse structures. We suppose that Groups~I and II roughly
represents young and more evloved PNs, respectively. There are 62 Groups~I PNs and 22 Group~II
PNs. We derive $[5.8]-[8.0]=1.52\pm0.57$ for Group~I and $1.47\pm0.33$ for Group~II, in reasonable
agreement with the conclusions of \citet{cohen11}. However, giving such a large standard deviation,
it is not possible to gain any further meaningful conclusion.

\subsection{M~1-41}

M~1-41 is one of the most interesting PNs within the GLIMPSE~3D survey area.  Its central star has a energy-balance temperature of 142\,400\,K \citep{pre91}, and is at a distance of about 1\,kpc \citep{zhang95}. Based on the radio morphology and infrared color, \citet{zij90} suggested that this source is a mis-classified PN and is likely to be a \ion{H}{2} region. However, a different conclusion was drawn by \citet{boh01}, who detected extended shock excited H$_2$ emission, and suggested that M~1-41 is a type~I PN.

\begin{figure*}
\epsfig{file=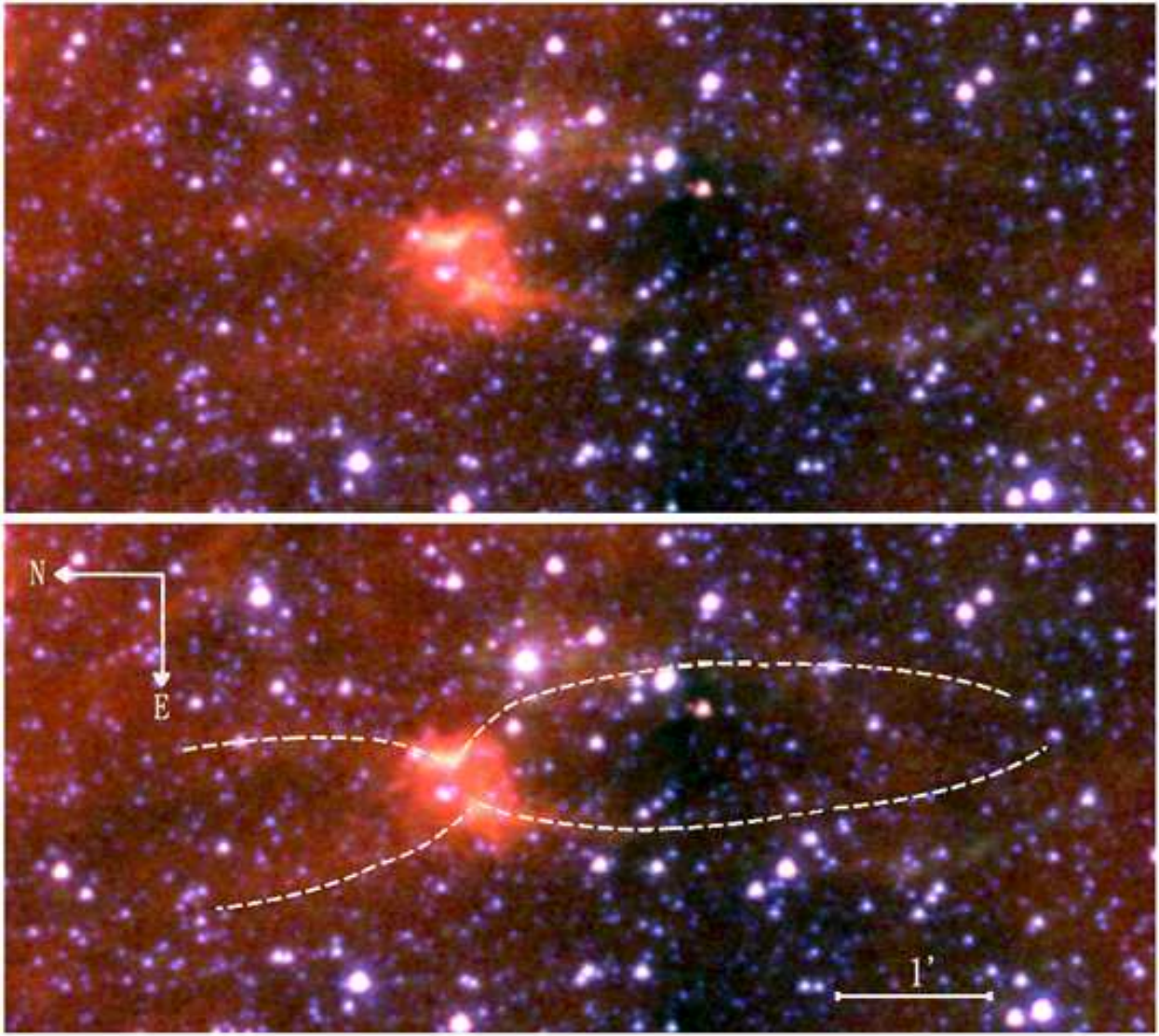, height=15cm}
\caption{The IRAC image of M~1-41. The 3.6\,$\mu$m is shown as blue, the 5.8\,$\mu$m is
shown as green, and the 8.0\,$\mu$m is shown as red. The superimposed dashed lines are a sketch of
the extended bipolar structures.\label{m141}}
\end{figure*}

Figure~\ref{m141} gives the IRAC image of M~1-41, which clearly reveals that this source is composed of a relatively bright central region and a pair of very extended faint lobes, suggesting that it is likely a nearby PN. The waist of the lobes is bright and visible in all the four bands. It is oriented at PA$=122^\circ$ and has a size of $\sim0.5'$. The lobes are visible only at 8\,$\mu$m. Note that the photometry data given in Table~1-3 are only based on the central part of this PN.
The northern lobe is incomplete because of the contamination from bright infrared background emission. The southern lobe has an extension of about $3.7'$ from the center. The long axis of the lobes is oriented at PA$=9^\circ$ and is not perpendicular to the waist. 
It is clear that if the instrumental sensitivity is not sufficient to detect the extended lobes, the irregular central nebulosity would appear to be a \ion{H}{2} regions, not a PN. This is a good example showing that poor dynamic range imaging can lead to mis-classification of the morphology of PNs.

Figure~\ref{sed} shows that  M~1-41 has a typical SED of PNs.  The excess in the near infrared suggests that there is a warm dust component in addition to the main cool dust component peaking at 30 $\mu$m.

The appearance of M~1-41 suggests that the mass of the central part is much larger than that of lobes. The bipolar lobes only manifest themselves through their thin walls, outlining very low density cavities. The walls of the lobes can be the result of sweeping up of previously-ejected circumstellar materials by a later developed, fast, collimated wind.
An alternative interpretation is that the bipolar cavities are created by radiation pressure blown out of the polar regions of an optically thick equatorial torus.
In this scenario, the bipolar structure is not caused by the dynamical
ejection, but by illumination. A similar scenario has been suggested to explain the formation of multipolar lobes of PNs \citep{kwok10,gar10}.

\begin{figure*}
\epsfig{file=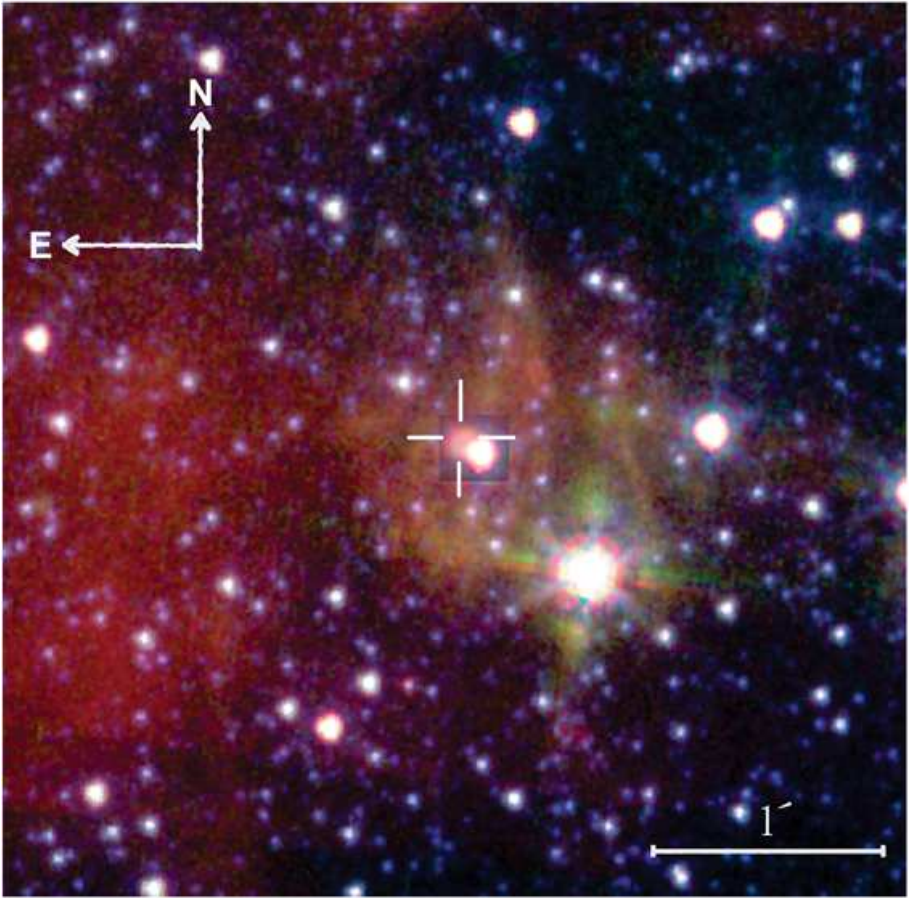, height=12cm}
\caption{The IRAC image of PN 1824$-$1410. The 3.6\,$\mu$m is shown as blue, the 5.8\,$\mu$m is
shown as green, and the 8.0\,$\mu$m is shown as red. The position of the
central part is marked.
\label{1824}}
\end{figure*}

Except for M~1-41, we do not find PNs that obviously exhibit extremely extended structures.  Most of the PNs can be clearly distinguished from the large-scale backgound emission.  An exception is the bright IRAS source, PN 1824$-$1410. This object was first identified as a PN by \citet{van96} based on optical observations. Figure~\ref{1824} shows its IRAC image. The central part is a red and compact nebula, which is partially obscured by a foreground bright star. Its color and morphology are typical of the GLIMPSE~3D sample PNs.  The IRAC image also reveals a more extended irregular emission region surrounding this source (about 1\,arcmin in radius), at the west of which some filaments can be seen and are aligned approximately north-south.  These IR structures are much more obvious compared to those shown by the H$\alpha$  image of \citet{van96}. Further investigation is needed to determine whether the extended irregular nebulosity is associated with this PN.


\section{DISCUSSION}

\begin{figure*}
\epsfig{file=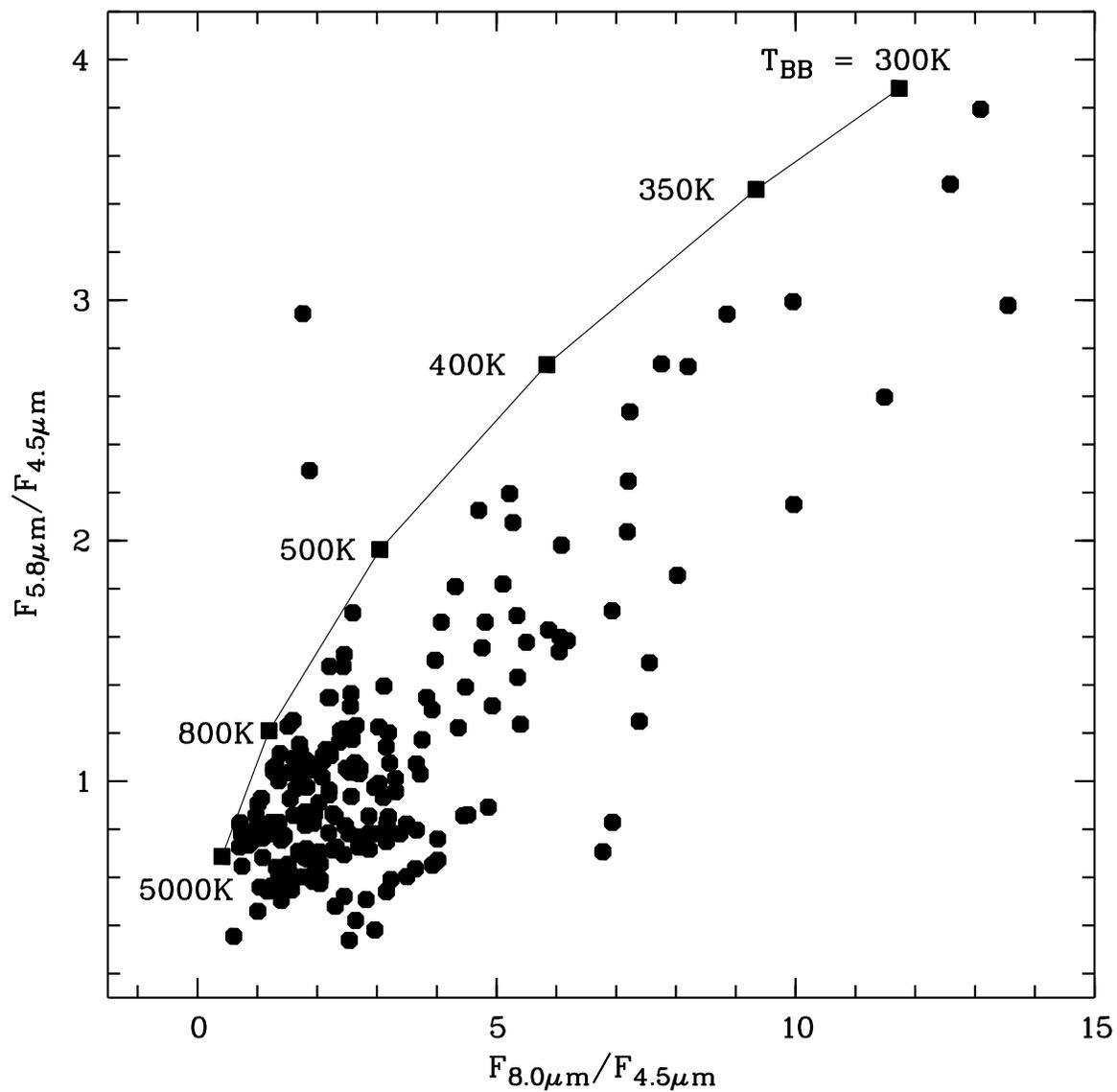, height=15cm}
\caption{The 5.8\,$\mu$m/4.5\,$\mu$m flux ratios vs  the 8.0\,$\mu$m/4.5\,$\mu$m flux ratios of 182 PNs in our sample.  Nine objects are not plotted due to non-detection or saturated fluxes in one or more of the bands.
The solid line is a track of blackbodies.  Some blackbody temperatures are marked on the curve.  
\label{corr}}
\end{figure*}

In order to investigate the contributions of AIBs to the IRAC bands we examine the relation between the 5.8$\mu$m/4.5$\mu$m and 8.0$\mu$m/4.5$\mu$m flux ratios
for the GLIMPSE~3D PNs (Figure~\ref{corr}).  It is clear that a positive correlation exists, and the distribution of the objects does not follow the blackbody curve.
The AIBs at 6.2 and 7.7\,$\mu$m
can contribute to the 5.8 and  8.0\,$\mu$m bands, respectively. As the 4.5 $\mu$m band has no contribution from AIBs, the positive correlation might reflect the correlation between the 6.2 and 7.7\,$\mu$m AIB strengths.
Figure~\ref{corr} also suggests that the contributions from AIB emission to the  8.0\,$\mu$m band is stronger than  that to the 5.8\,$\mu$m band.  
Another factor that contributes to the observed deviation to the blackbody curve is the contamination from the emission of ionized gas 
and/or central star to the 4.5\,$\mu$m band.  
This point can be verified in Figure~\ref{image} where we do find that the 4.5\,$\mu$m images are generally more 
centrally enhanced than the 5.8 and  8.0\,$\mu$m ones. The effect can cause the data to shift towards higher temperature regions in this plot, and is more
pronounced for the blackbody temperatures estimated by the 5.8$\mu$m/4.5$\mu$m flux ratio.

\begin{figure*}
\epsfig{file=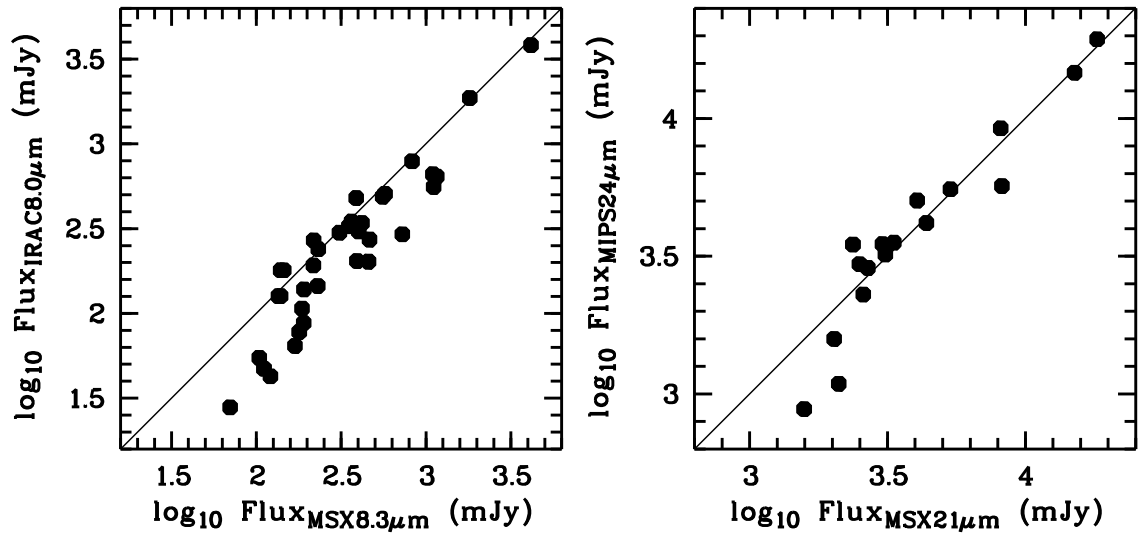, height=7cm}
\caption{Left panel: IRAC 8.0\,$\mu$m vs. {\it MSX} 8.3\,$\mu$m integrated fluxes for 35 PNs. Right panel: MIPS 24\,$\mu$m vs. {\it MSX} 21\,$\mu$m integrated fluxes for 17 PNs.
The solid line is a $y=x$ plot.
\label{cal}}
\end{figure*}

In order to examine the reliability of our flux measurements and the flux calibration of extended sources,  we compare the  IRAC 8.0\,$\mu$m vs. {\it MSX} 8.3\,$\mu$m and the MIPS 24\,$\mu$m vs. {\it MSX} 21\,$\mu$m integrated fluxes of The GLIMPSE~3D PNs in Figure~\ref{cal}. 
For these compact extended sources, the IRAC 8.0\,$\mu$m and MIPS 24\,$\mu$m
fluxes are in good agreement with those of {\it MSX} 8.3\,$\mu$m and 21\,$\mu$m, respectively. We obtain average IRAC8.0\,$\mu$m/{\it MSX}8.3\,$\mu$m and MIPS24\,$\mu$m/{\it MSX}21\,$\mu$m flux ratios of $0.74\pm0.28$ and $0.99\pm0.24$, respectively. 
This is consistent with our previous results based on the GLIMPSE~II PNs \citep{zhang09}, and suggests  that our flux measurements are reliable. The average IRAC8.0\,$\mu$m/{\it MSX}8.3\,$\mu$m ratio also agrees with that of PNs obtained by \citet{cohen07b}, but lower than that of \ion{H}{2} regions deduced by \citet{cohen07a} ($1.55\pm0.15$), 
suggesting that lower aperture correction factor should be applied to
obtain the IRAC fluxes of more extended source. Figure~~\ref{cal}
also indicates that for faint PNs the {\it MSX} 8.3\,$\mu$m and
21\,$\mu$m fluxes tend to be underestimated. This might be due to
the lower instrumental sensitivity of {\it MSX}.

\begin{figure*}
\epsfig{file=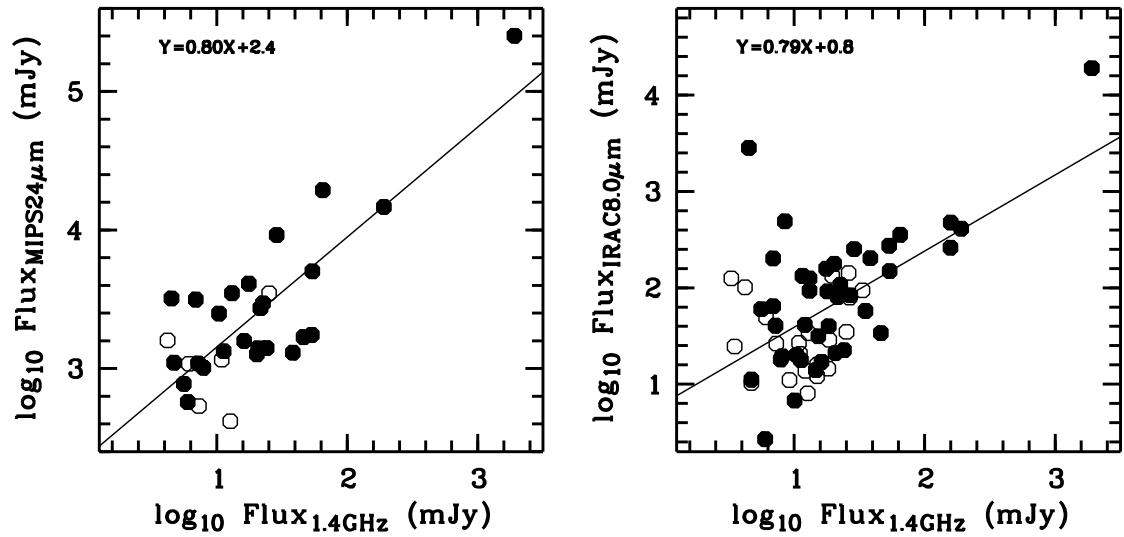, height=7cm}
\caption{Left panel: MIPS 24\,$\mu$m vs. NVSS 1.4\,GHz integrated fluxes for 31 PNs.
Right panel: IRAC 8.0\,$\mu$m vs. NVSS 1.4\,GHz integrated fluxes for 61 PNs.
The solid line represents a linear fitting. The open and filled circles
denote the MASH and known PNs, respectively.
\label{radio}}
\end{figure*}

As PNs expand and disperse into the interstellar medium, both the radio and infrared fluxes are expected to decline with age.  It is therefore useful to see if a correlation exists between these two fluxes.
\citet{cohen07b} claimed that previously known PNs and more evolved
MASH PNs have different IR/radio flux ratio. Their conclusion, however,
is not supported by subsequent studies \citep{cohen11,phillips11}.
In Figure~\ref{radio} we compare the IR fluxes at 8.0\,$\mu$m
and 24\,$\mu$m and the radio flux at 1.4\,GHz from NVSS \citep{condon98}.
There is no systematically difference between the IR/radio
fluxes of MASH PNs and those of previously known PNs although MASH PNs
are generally fainter. 
Figure~\ref{radio} exhibits a weak correlation between the IR and radio 
fluxes. The distribution of objects in the 8.0\,$\mu$m vs. 1.4\,GHz flux plot
are more scattered than that in the 24\,$\mu$m and vs. 1.4\,GHz plot, which might be attributed to the contamination from AIBs to the 8.0\,$\mu$m band.

\citet{phillips09} compared the IR colors of Galactic PNs and those
of Large Magellanic Could (LMC) PNs and found that the LMC PNs have 
lower [5.8]-[8.0] color indices.  However, based on a study of different
sample \citet{cohen11} argued that there is no statistically meaningful 
difference between these IR colors. Comparing the color-color plot 
(Figure~\ref{color}) with Figure~6 of \citet{phillips09}, we find
that the color indices of GLIMPSE~3D PNs are approximately located within
the same range with those of LMC PNs. The average [5.8]-[8.0] color
indices of GLIMPSE~3D PNs is 1.62 which is not much different from
the average value of the LMC PNs. Our results, therefore, support the 
conclusion of \citet{cohen11}.

\section{CONCLUSIONS}

From the GLIMPSE 3D survey data, we have identified the IR counterparts of 191 galactic PNs.  The images of 90 MASH PNs and 80 previous known PNs are presented and the SEDs of 83 PNs are constructed.  The SEDs show clearly the importance of the dust component in PNs, as in many objects most of the energy is emitted in the dust component.  The set of PN SEDs presented in this paper has helped us define the observational properties of PNs, allowing us to distinguish PN from other emission line objects. 

Very extended bipolar lobes are discovered in the PN M1-41, suggesting that IR imaging is useful in finding outer structures of PNs which may be missed in optical observations.  

One of the conclusions we have from this study is that the infrared images of the PNs are somewhat different from those in the visible.  The obvious explanation is that the dust is distributed differently from the ionized gas region.  For example, bipolar nebulae would be visibly bright in the lobes but infrared bright in the equatorial regions.
Interstellar extinction may also have affected the optical appearance of the objects.

The GLIMPSE IR data are useful to search for new PNs. The IRAC observations have resulted in the discovery of 
a new extremely redden PN G313.3+00.3 which is optically invisible \citep{cohen05a}.
This suggests that there may be many PNs hidden in the Galactic plane and the current census of Galactic PNs is far from being complete. The number of detected Galactic PNs is about one order of magnitude lower than the theoretically predicted value. We are starting a project to search for new (optically invisible) PNs in the GLIMPSE field.  Our results will be reported elsewhere. The IR colors and SEDs of PNs presented in this paper will provide a useful diagnostics for the identification of new PNs.


\acknowledgments

We are grateful to Ed Churchwell and the GLIMPSE team for assistance in the processing and analysis of GLIMPSE survey data.  We also thank Nico Koning for helpful discussions.  This work is based on observations made with the {\it Spitzer Space Telescope}, which is operated by the Jet Propulsion Laboratory, California Institute of Technology, under a contract with NASA.  This research has also made use of the NASA/IPAC Infrared Science Archive, which is operated by the Jet Propulsion Laboratory, California Institute of Technology, under contract with the National Aeronautics and Space 
Administration.  Support for this work was provided by the Research Grants Council of the Hong Kong under grants HKU7032/09P and the Seed Funding Programme for Basic Research in HKU (200909159007).

\appendix

\section{INDIVIDUAL MASH PNs}

 {PNG $000.4+04.4$ }.--- 
This object, previously known as PN K~5-1, is a compact and low-excitation PN. 
It was first assigned as a possible PN by \citet{pre88} from the IRAS point Source Catalogue.
The IRAC image shows that its IR emission is more extended and diffuse than its H$\alpha$
counterpart (note that the optical images hereafter
are based on the MASH catalogue). This might suggest that there exists
extended dust and the nebula is ionized bounded. The SED indicates
a color temperature of $T<100$\,K.

 {PNG $001.0+02.2$ }.---
Its PN assignation can be confirmed by the IRAC color. The IR appearance is clearly more extended than its optical counterpart. The SED indicates 
a color temperature of $T<150$\,K.

 {PNG $001.1+02.2$ }.---
The IRAC image shows that it has a compact and bright core. We cannot
find any difference between its IR and optical appearances. Its SED
seems trace a component with a color temperature of about 600\,K.

 {PNG $001.2+02.8$ }.---
The low IR fluxes and compact appearance suggest that it is distant.

 {PNG $001.5+03.1$ }.---
The IRAC image shows that it is diffuse and has an oval shape.
The SED suggests that it is surrounded by dust
with a temperature of about 100\,K.

 {PNG $001.5-02.8$ }.---
The IRAC image clearly shows that it consists of a bright core and an
oval nebula. The nebula has a well-defined boundary and seems to be
fainter along its major axis direction.

 {PNG $001.7+03.6$ }.---
The H$\alpha$ image shows that it  is bright and compact. However,
its IR counterpart is almost overwhelmed by the bright background
emission.

 {PNG $001.7-02.6$ }.---
It has weak hydrogen emission lines. The H$\alpha$ image shows
that ``it is a very small faint round PN with enhanced E-W
limbs/condensations''. The limbs/condensations are not clear in
the IRAC image.

 {PNG $002.1+02.6$ }.---
The IRAC image shows that  it is faint and has an oval shape.

 {PNG $002.2+01.7$ }.---
Both H$\alpha$ and IRAC images reveal that it is extremely compact.

 {PNG $002.3-01.7$ }.---
It is a compact object. \citet{parker06} suggests that it might be
an emission line star. Its IR emission is strong.  The SED shows
that it has a dust component with a color temperature of $T<100$\,K
and is very likely to be a PN.

 {PNG $002.3+01.7$ }.---
It has an oval shape.  The H$\alpha$ image shows that it has E-W limbs, which however is not clear in the IRAC image.

 {PNG $002.5+02.0$ }.---
The H$\alpha$ and IRAC images show that it is a faint compact source.
The SED indicate to a color temperature of $T<150$\,K.

 {PNG $002.7+01.7$ }.---
The H$\alpha$ and IRAC images reveal a round appearance.

 {PNG $002.9-02.7$ }.---
This source is also known as K\,6-39 \citep{koh02}. 
It is compact and has a high excitation spectrum.
The SED indicates a strong nebular bound-free emission and a
dust component with a temperature of about 300\,K.

 {PNG $003.1-01.6$ }.---
The H$\alpha$ image shows that it is an oval PN with 
faint outer detached halo to north. The IRAC image reveals a
bright core with faint extended nebulosity.

 {PNG $003.4-01.8$ }.---
The H$\alpha$ image shows that it is a compact bipolar nebula.
The bipolar structure, however, is not clear in the IRAC image.

 {PNG $004.2-02.5$ }.---
The H$\alpha$ and IRAC images reveal an oval shape.

 {PNG $004.3+01.8a$}.---
It has an oval shape.  The IRAC image shows a diffuse structure.

 {PNG $005.2-01.6$ }.---
Both H$\alpha$ and IRAC images show a very compact structure.

 {PNG $005.8+02.2a$}.---
Its H$\alpha$ appearance looks like a fuzzy elliptical nebula. The IRAC image
reveal a bright core and a possible bipolar structure elongated along the
NW-SE direction. The SED indicates a color temperature of $<300$\,K.

 {PNG $006.0+01.2$ }.---
The IRAC image reveals an fuzzy oval shape. However, its 24\,$\mu$m emission
is quite strong.

 {PNG $006.1+01.5$ }.---
This source is also known as K\,6-33 \citep{koh02}. 
Both H$\alpha$ and IRAC images show an X-shaped structure,
suggesting that it might be a bipolar PN. The SED indicates 
a color temperature of about 100\,K

 {PNG $006.1-02.1$ }.---
Both H$\alpha$ and IRAC images reveal a ring morphology.

 {PNG $006.9+01.5$ }.---
The IRAC image shows that it is an extremely compact nebula.

 {PNG $007.5-02.4$ }.---
The H$\alpha$ image reveals an approximately circular nebula. 
The IRAC image shows a compact oval shape. The SED  suggests a
cold dust component with $T<100$\,K.

 {PNG $007.8+01.2$ }.---
It is faint in optical image, but clearly seen by IRAC. It has a
compact round morphology.

 {PNG $009.4-01.2$ }.---
It shows a slightly oval shape. The IRAC appearance is very diffuse, and
almost overwhelmed by the background emission. The SED shows that it has
strong far-IR emission.

 {PNG $009.7-01.1$ }.---
The H$\alpha$ image shows a faint oval shape. It can be more clearly
viem in the IRAC image.
The SED suggests that its IR emission is dominated by thermal emission
from cold dust with a temperature of $<50$\,K. This object may suffer from
heavy extinction.

 {PNG $009.8-01.1$ }.---
Both H$\alpha$ and IRAC images reveal a bipolar structure with
a bright core. Its IRAC appearance looks more centrally enhanced, suggesting
the presence of a dust torus.

 {PNG $010.0-01.5$ }.---
The H$\alpha$ and IRAC images reveal a compace oval structure.
The SED suggests a color temperature of $<$300\,K.

 {PNG $010.2+02.4$ }.---
The IRAC image reveals a roughly oval shape with some irregular filaments.

 {PNG $010.2+02.7$ }.---
This source is also known as IRAS 17552-1841 \citep{koh01}.
The  H$\alpha$ and IRAC images reveal a circular morphology.
The SED suggests  strong far-infrared emission, implying the presence
of cold dust.

 {PNG $010.6+02.4$ }.---
The H$\alpha$  image shows a round shape. The IRAC image reveals a slightly
oval structure.

 {PNG $010.7-02.3$ }.---
The H$\alpha$ and IRAC images reveal a vaguely oval structure with
a tail located on its southwest side.

 {PNG $011.0+01.4$ }.---
The IRAC image reveals an oval structure with diffuse end of the major
axis. The H$\alpha$ shows the background emission, which is brighter
in the IRAC image.

 {PNG $011.0-02.9$ }.---
It appears a round shape. Its PN status was further confirmed
by \citet{bou06}. The SED indicates the presence of cold dust
with $T<100$\,K.

 {PNG $017.6+02.6$ }.---
Its H$\alpha$ appearance is compact. The IRAC image shows an
oval shape with well-defined boundary. The SED reveals strong
IR emission with a color temperature of $<100$\,K.

 {PNG $019.2-01.6$ }.---
The H$\alpha$ and IRAC images show a round shape.
The SED indicates a color temperature of $<300$\,K. 

 {PNG $024.2+01.8$ }.---
The H$\alpha$ and IRAC images show a roughly round structure with
diffuse boundary.

 {PNG $025.6+02.8$ }.---
The IRAC image reveal a bright compact core.
The SED indicates a color temperature of about $150$\,K.

 {PNG $029.0+02.2$ }.---
It is unresolved in the IRAC image.
The SED indicates a color temperature of $<100$\,K.

 {PNG $029.2-01.8$ }.---
The H$\alpha$ image reveals a circular shape. In the IRAC image, it
is heavily obscured by the bright background emission.

 {PNG $029.4-02.3$ }.---
The IRAC image reveals a slightly oval shape although its H$\alpha$
image appears to be round.

 {PNG $030.2+01.5$ }.---
It is a compact object.

 {PNG $031.0-02.1$ }.---
Both H$\alpha$ and IRAC images reveal an elliptical shape.

 {PNG $329.8-03.0$ }.---
The H$\alpha$ image shows that it is a compact, bright oval nebula.
The IRAC image reveal a faint ring  with some irregular structures
in the boundary.  The SED suggests strong far-IR emission from
cold dust.

 {PNG $330.1+02.6$ }.---
It is a compact source.  The IRAC image reveals a bright core with a faint halo.
The SED suggests a color temperature of about 150\,K.

 {PNG $330.7+02.7$ }.---
Its H$\alpha$ emission is very faint. The IRAC image reveals a bright 
compact source.

 {PNG $334.0+02.4$ }.---
The H$\alpha$ and IRAC images exhibit an elliptical shape.
The SED indicates the presence of cold dust with a color temperature of 60--150\,K.

 {PNG $334.4+02.3$ }.---
The H$\alpha$ and IRAC images show an elliptical shape.
The SED suggests a color temperature of about 150\,K.

 {PNG $335.4-01.9$ }.---
The H$\alpha$ and IRAC images show an extended elliptical shape. 
The SED indicates the presence of cold dust with a color temperature of 60--150\,K.

 {PNG $335.8-01.6$ }.---
The IRAC image exhibits a bright compact source. Its SED is similar to those
of other PNs, and suggests the presence of cold dust with a temperature of $<150$\,K.

 {PNG $335.9-01.3$ }.---
The IRAC image reveal a bright compact source.
Its SED clearly shows the nebular bound-free emssion, and dust emission
with a color temperature of about 150\,K.

 {PNG $341.7+02.6$ }.---
It has an annular shape. The IRAC image shows that it is located inside
a large cloud.
The SED suggests a color temperature of $<300$\,K.

 {PNG $341.9-02.8$ }.---
The IRAC image shows a compact source.
The SED reveals a sharp increase in the wavelength of about 10\,$\mu$m.

 {PNG $342.1-02.0$ }.---
It is a very compact object.

 {PNG $344.0+02.5$ }.---
The H$\alpha$ and IRAC image reveal a compact structure.
The SED suggests a color temperature of $<150$\,K.

 {PNG $344.4+01.8$ }.---
The H$\alpha$ image shows an elliptical morphology. The IRAC image
reveals some filaments around the elliptical structure.
The SED suggests a color temperature of $<300$\,K.

 {PNG $344.8-02.6$ }.---
The H$\alpha$ and IRAC image reveal a compact appearance.

 {PNG $345.8+02.4$ }.---
The H$\alpha$ and IRAC image reveal a compact structure.
The SED suggests a color temperature of $<300$\,K.

 {PNG $345.8+02.7$ }.---
Both H$\alpha$ and IRAC images exhibit a oval ring with outer extensions.
The central star as revealed in the H$\alpha$ image is invisible in
the IRAC image. The SED suggests a color temperature of $<300$\,K.

 {PNG $349.6-02.1$ }.---
The IRAC image shows a fuzzy structure with bright background emission.

 {PNG $350.4+02.0$ }.---
This source is also known as IRAS 17092-3539 \citep{koh01}.  The H$\alpha$ and IRAC image reveal a faint elongated
shape.  The SED suggests a color temperature of $\sim100$\,K.

 {PNG $350.8+01.7$ }.---
This source is also known as  IRAS 17114-3532 \citep{koh01}.
The H$\alpha$ and IRAC image reveal a round appearance. The SED suggests the
presence of cold dust with a temperature of $\sim100$\,K.

 {PNG $350.8-03.0$ }.---
The IRAC image shows an extremely compact structure.

 {PNG $355.0+02.6$ }.---
This source is also known as  IRAS 17194-3137 \citep{koh01}.
The IRAC image reveals a round appearance. The SED suggests a color temperature of $\sim100$\,K.

 {PNG $355.2-02.0$ }.---
The H$\alpha$ and IRAC image reveal a round appearance.

 {PNG $355.8+01.7$ }.---
It is a very compact object.

 {PNG $356.0-01.8$ }.---
The H$\alpha$ and IRAC image reveal a compact and round appearance.

 {PNG $356.0+02.8$ }.---
This source is also known as  IRAS 17217-3040 \citep{koh01}.
The H$\alpha$ and IRAC image reveal a compact structure. 
The SED suggests a color temperature of $<150$\,K.

 {PNG $356.1-02.7$ }.---
It is a compact object. The IRAC appearance looks more extended than
the optical one.

 {PNG $356.2+02.5$ }.---
The H$\alpha$ and IRAC image reveal a compact structure.
The SED suggests that its far-IR emission is quite strong.

 {PNG $356.2+02.7$ }.---
The H$\alpha$ and IRAC image reveal a compact structure.

 {PNG $356.3-02.6$ }.---
It is a compact nebulosity behind a field star.

 {PNG $356.5-01.8$ }.---
It is compact object.
The SED may indicate the presence of a warm component, and thus
 it might be symbiotic star, as suggested by \citet{parker06}.

 {PNG $356.6+02.3$ }.---
The optical and IRAC images reveal a ring structure. Faint extentions
are also displayed by the IRAC image.

 {PNG $357.3-02.0$ }.---
The H$\alpha$ image shows that it is a faint PN in obscured region.
It appears as a bright round nebulosity in the IRAC image.
The SED indicates a color temperature of $\sim100$\,K.

 {PNG $357.5-02.4$ }.---
The H$\alpha$ and IRAC images reveal a faint oval structure.
The SED indicates a color temperature of $<100$\,K.

 {PNG $357.8+01.6$ }.---
The H$\alpha$ and IRAC images reveal a slightly extended structure.
The H$\alpha$ image also displays an outer arm, which however is not
clear in the IRAC image.

 {PNG $357.9+01.7$ }.---
It is a compact object.

 {PNG $358.0-02.4$ }.---
The H$\alpha$ and IRAC images show a faint compact structure.

 {PNG $358.1+02.3$ }.---
The IRAC image reveals a bright core with slightly extended structure.
Its SED differs from those of other PNs and
suggests a color temperature of $\sim600$\,K. Thus it is unlikely to be
a PN.

 {PNG $358.4+02.1$ }.---
Its IR appearance is very fuzzy and almost overwhelmed by the
background emission.

 {PNG $358.7-02.5$ }.---
This source is also known as K\,6-31 \citep{koh02}.
The IRAC image shows a faint compact nebulosity.

 {PNG $359.2-02.4$ }.---
The IRAC image shows a bright core with oval extentents.

 {PNG $359.4+02.3a$}.---
The H$\alpha$ and IRAC images show a compact structure.

 {PNG $359.6+04.3$ }.---
The IRAC image exhibits a faint small nebulosity.

 {PNG $359.7+02.0$ }.---
It is a compact object.
The SED suggests a color temperature of $<150$\,K.

 {PNG $359.8+03.5$ }.---
The H$\alpha$ and IRAC images reveal an oval nebulosity.

\clearpage

\clearpage

\clearpage

\clearpage

\clearpage

\clearpage

\clearpage

\clearpage

\begin{deluxetable}{lcc@{\extracolsep{0.1in}}ccccc}
\tabletypesize{\scriptsize} 
\tablecaption{Spitzer observations of the GLIMPSE~3D PNs. \label{flux}} \tablewidth{0pt}
\tablehead{
Objects$^a$ & \multicolumn{2}{c}{Coordinate (J2000.0)} & \multicolumn{5}{c}{Flux (mJy)$^b$} \\
\cline{2-3}\cline{4-8} 
 & \colhead{R.A.} & \colhead{Dec.} &  \colhead{3.6$\micron$} & \colhead{4.5$\micron$} & \colhead{5.8$\micron$} & 
\colhead{8.0$\micron$} & \colhead{24$\micron$}} 
\startdata
{\bf MASH PNs}\\
PNG $000.4+04.4$ I t& 17 29 52.4 & -26 11 13 & 3.05 & 5.83 & 3.65 & 20.66 & \nodata \\
PNG $001.0+02.2$ I l& 17 39 11.3 & -26 52 20 & 7.24 & 8.42 & 10.67 & 18.93 & 210.98 \\ 
PNG $001.1+02.2$ II t& 17 39 49.7 & -26 48 45 & 7.24 & 9.70 & 11.08 & 12.91 & \nodata \\
PNG $001.2+02.8$ I l& 17 37 30.3 & -26 21 44 & 7.07 & 5.94 & 5.94 & 13.71 & 60.96 \\
PNG $001.5+03.1$ I t& 17 37 16.5 & -25 59 38 & 8.80 & 10.94 & 11.92 & 24.66 & \nodata \\
PNG $001.5-02.8$ I p& 18 00 22.4 & -29 04 39 & 6.97 & 8.54 & 13.50 & 19.44 & \nodata \\ 
PNG $001.7+03.6$ I t& 17 35 47.4 & -25 27 43 & 7.05 & 6.29 & 5.64 & 12.18 & 325.37 \\
PNG $001.7-02.6$ I t& 18 00 00.6 & -28 46 27 & 4.87 & 4.33 & 2.75 & 6.81 & 614.37 \\
PNG $002.1+02.6$ I t& 17 40 30.7 & -25 44 40 & 13.00 & 9.03 & 7.60 & 8.00 & 148.69 \\
PNG $002.2+01.7$ I p& 17 44 01.7 & -26 05 45 & 5.13 & 6.76 & 5.09 & 10.92 & 139.36 \\
PNG $002.3+01.7$ I t& 17 44 35.4 & -26 03 36 & 5.44 & 5.70 & 6.41 & 12.06 & 79.72 \\
PNG $002.3-01.7$ I p& 17 57 43.5 & -27 50 44 & 25.22 & 26.17 & 20.87 & 102.57 & 881.18 \\
PNG $002.5+02.0$ I l& 17 43 39.5 & -25 38 17 & 6.71 & 8.78 & 8.87 & 16.48 & 150.33 \\
PNG $002.7+01.7$ I t& 17 45 18.9 & -25 42 05 & 18.97 & 12.50 & 8.46 & 7.78 & 72.56 \\
PNG $002.9-02.7$ I l& 18 03 05.1 & -27 46 44 & 25.00 & 24.81 & 41.73 & 93.72 & 357.81 \\
PNG $003.1-01.6$ I t& 17 59 26.1 & -27 06 34 & 56.14 & 44.92 & 48.07 & 97.62 &  345.64 \\
PNG $003.4-01.8$ I t& 18 00 42.3 & -26 53 37 & 11.36 & 9.14 & 7.39 & 14.35 & 115.73 \\
PNG $004.2-02.5$ I t& 18 05 20.1 & -26 31 45 & 6.49 & 8.23 & 4.46 & 13.96 & 429.43 \\
PNG $004.3+01.8a$ I t& 17 48 33.0 & -24 17 36 & 5.27 & 5.41 & 4.67 & 10.17 & 367.37 \\
PNG $005.2-01.6$ I t& 18 03 52.4 & -25 16 59 & \nodata & 27.60 & \nodata & 39.39 & 162.06 \\
PNG $005.8+02.2a$ II t& 17 50 20.7 & -22 48 24  & 12.29 & 16.55 & 23.97 & 65.31 & \nodata \\
PNG $006.0+01.2$ I p& 17 54 44.8 & -23 09 07 & 13.05 & 12.86 & 10.17 & 10.89 & 500.87 \\
PNG $006.1+01.5$ I t& 17 53 45.3 & -22 54 01 & 11.03 & 22.42 & 14.77 & 34.89 & 3487.26 \\
PNG $006.1-02.1$ I t& 18 07 40.9 & -24 39 18 & 1.56 & 2.73 & 3.50 & 10.57 & 225.94 \\
PNG $006.9+01.5$ II t& 17 55 36.7 & -22 12 47  & 5.04 & 6.19 & 4.28 & 5.02 & \nodata \\
PNG $007.5-02.4$ I t& 18 11 40.6 & -23 37 16 & 4.57 & 9.57 & 5.03 & 11.03 & \nodata \\
PNG $007.8+01.2$ II l& 17 58 33.9 & -21 35 15  & 2.85 & 4.23 & 2.66 & 6.84 & \nodata \\
PNG $009.4-01.2$ I t& 18 11 10.6 & -21 23 14 & 12.77 & 13.86 & 13.02 & 28.76 & \nodata \\
PNG $009.7-01.1$ I p& 18 11 41.3 & -21 02 29 & 1.30 & 2.02 & 5.61 & 24.15 & \nodata \\
PNG $009.8-01.1$ I t& 18 11 39.0 & -21 00 44 & 13.81 & 27.93 & 56.37 & 125.31 & \nodata \\ 
PNG $010.0-01.5$ I t& 18 13 35.4 & -20 57 05 & 24.17 & 42.96 & 60.90 & 93.89 & \nodata \\
PNG $010.2+02.4$ I l& 17 59 05.1 & -18 53 22 & 3.85 & 4.47 & 4.32 & 7.25 & \nodata \\
PNG $010.2+02.7$ I t& 17 58 14.4 & -18 41 26 & 3.33 & 12.20 & 6.82 & 14.45 & \nodata \\
PNG $010.6+02.4$ II t& 18 00 08.2 & -18 34 34  & 4.64 & 4.64 & 5.82 & 9.07 & \nodata \\
PNG $010.7-02.3$ II t& 18 18 04.4 & -20 44 14  & 11.56 & 15.48 & 9.18 & 17.66 & \nodata \\
PNG $011.0+01.4$ I t& 18 04 29.2 & -18 42 41 & 26.03 & 23.77 & 24.40 & 33.77 & \nodata \\
PNG $011.0-02.9$ I t& 18 20 53.8 & -20 48 12 & 6.72 & 9.02 & 5.89 & 15.96 & \nodata \\
PNG $017.6+02.6$ I t& 18 13 33.7 & -12 20 48 & 6.86 & 13.51 & 8.80 & 79.05 & \nodata \\
PNG $019.2-01.6$ II t& 18 31 53.0 & -12 56 14  & 4.30 & 8.79 & 4.48 & 12.11 & \nodata \\
PNG $024.2+01.8$ II t& 18 28 48.0 & -06 52 00  & 1.88 & 2.72 & 1.80 & 6.78 & \nodata \\
PNG $025.6+02.8$ II t& 18 27 57.4 & -05 10 23  & 2.15 & 3.17 & 2.35 & 10.27 & \nodata \\
PNG $029.0+02.2$ II t& 18 36 10.7 & -02 27 14  & 4.41 & 8.82 & 4.94 & 13.73 & \nodata \\
PNG $029.2-01.8$ I t& 18 51 12.1 & -04 08 55 & 2.93 & 3.83 & 4.33 & 8.23: & \nodata \\
PNG $029.4-02.3$ II t& 18 53 30.6 & -04 14 09  & 4.35 & 3.53 & 1.66 & 4.38 & \nodata \\
PNG $030.2+01.5$ II t& 18 41 04.5 & -01 40 49  & 2.28 & 4.18 & 4.30 & 7.72 & \nodata \\
PNG $031.0-02.1$ II t& 18 55 34.6 &  02 40 27  & 4.11 & 4.16 & 3.94 & 7.60 & \nodata \\
PNG $329.8-03.0$ I t & 16 17 19.8 & -54 45 12 & 3.96 & \nodata & 9.76 & \nodata & \nodata \\
PNG $330.1+02.6$ II t& 15 54 08.9 & -50 22 39  & 6.66 & 8.07 & 12.22 & 41.53 & \nodata \\
PNG $330.7+02.7$ I t & 15 56 33.0 & -49 55 55 & 86.96 & 94.47 & 90.62 & 107.24 & \nodata \\
PNG $334.0+02.4$ II t& 16 12 40.6 & -47 50 14  & 3.61 & 4.89 & 5.89 & 16.80 & \nodata \\
PNG $334.4+02.3$ II t& 16 14 52.3 & -47 42 11  & 3.61 & 9.11 & 4.73 & 11.00 & \nodata \\
PNG $335.4-01.9$ I t& 16 37 44.9 & -49 57 50 & 14.51 & 19.49 & 13.91 & 47.59 & \nodata \\
PNG $335.8-01.6$ II p& 16 38 01.7 & -49 27 18  & 11.69 & 17.41 & 61.41 & 199.82 & \nodata \\
PNG $335.9-01.3$ II p& 16 37 14.2 & -49 11 16  & 20.36 & 25.59 & 61.77 & 257.63 & \nodata \\
PNG $341.7+02.6$ I t& 16 42 18.6 & -42 14 45 & 6.51 & 9.40 & 18.16 & 43.94: & \nodata \\
PNG $341.9-02.8$ II t& 17 06 25.2 & -45 27 04  & 2.78 & 5.11 & 2.99 & 8.86 & \nodata \\
PNG $342.1-02.0$ II t& 17 03 24.0 & -44 50 30  & 1.50 & 3.25 & 1.39 & 2.90 & \nodata \\
PNG $344.0+02.5$ II t& 16 50 20.2 & -40 30 03  & 1.84 & 2.83 & 2.46 & 7.75 & \nodata \\
PNG $344.4+01.8$ II l& 16 54 43.2 & -40 41 47  & 3.10 & 4.12 & 4.68 & 10.85 & \nodata \\
PNG $344.8-02.6$ II t& 17 15 15.9 & -43 03 54  & 3.55 & 5.17 & 3.73 & 5.91 & \nodata \\
PNG $345.8+02.4$ II t& 16 56 40.1 & -39 12 37  & 1.46 & 2.55 & 2.52 & 8.34 & \nodata \\
PNG $345.8+02.7$ I t& 16 55 51.9 & -39 00 21 & 1.83 & 2.63 & 5.16 & 10.68 & \nodata \\
PNG $349.6-02.1$ II t& 17 27 36.8 & -38 51 09  & 1.81 & 2.16 & 2.24 & 2.63 & \nodata \\
PNG $350.4+02.0$ I l& 17 12 34.0 & -35 43 20 & 12.97 & 13.54 & 10.73 & 33.82 & \nodata \\
PNG $350.8+01.7$ II p& 17 14 49.8 & -35 35 40  & 1.19 & 1.49 & 8.54 & 133.01 & \nodata \\
PNG $350.8-03.0$ I t& 17 34 52.8 & -38 17 19 & 1.59 & 1.11 & 1.03 & 3.28 & \nodata \\
PNG $355.0+02.6$ I t& 17 22 40.8 & -31 39 55 & 21.79 & 26.77 & 39.78 & 142.29 & \nodata \\
PNG $355.2-02.0$ I t& 17 41 59.1 & -34 05 34 & 27.14 & 34.66 & 35.32 & 49.19 & \nodata \\
PNG $355.8+01.7$ II t& 17 28 31.1 & -31 32 09  & 4.40 & \nodata & 7.17 & \nodata & \nodata \\
PNG $356.0+02.8$ I t& 17 24 58.3 & -30 43 04 & 5.94 & 7.50 & 9.13 & 17.09 & 652.38 \\
PNG $356.0-01.8$ I t& 17 43 07.2 & -33 15 54 & 4.21 & 6.83 & 3.81 & 9.33 & 556.13 \\
PNG $356.1-02.7$ I t& 17 47 04.8 & -33 41 03 & 1.14 & 2.72 & 1.97 & 8.19 & 451.52 \\
PNG $356.2+02.5$ I t& 17 26 23.6 & -30 45 39 & 4.89 & 5.78 & 4.94 & 8.01 & 416.23 \\
PNG $356.2+02.7$ II t& 17 25 33.4 & -30 33 57  & 1.67 & 2.04 & 2.15 & 3.89 & \nodata \\
PNG $356.3-02.6$ II t& 17 47 27.5 & -33 26 38  & 18.71 & 12.06 & 8.24 & 9.03 & \nodata \\
PNG $356.5-01.8$ I p& 17 44 28.0 & -32 52 11 & 31.66 & 31.63 & 30.72 & 49.38 & 1082.72 \\
PNG $356.6+02.3$ I t& 17 28 14.2 & -30 32 14 & 3.09 & 3.94 & 4.06 & 7.76 & 149.27 \\
PNG $357.3-02.0$ I t& 17 47 28.5 & -32 15 46 & 14.53 & 15.94 & 33.35 & 101.19 & 1593.93 \\
PNG $357.5-02.4$ I t& 17 49 37.9 & -32 16 28 & 7.22 & 9.47 & 4.76 & 26.35 & 535.03 \\
PNG $357.8+01.6$ I t& 17 34 01.7 & -29 54 35 & 47.99 & \nodata & 20.90 & \nodata & 620.67 \\
PNG $357.9+01.7$ I l& 17 33 38.4 & -29 45 30 & 2.27 & 3.46 & 3.74 & 11.21 & 420.71 \\
PNG $358.0-02.4$ I t& 17 50 48.5 & -31 52 27 & 13.79 & 10.62 & 9.56 & 15.57 & 535.93 \\
PNG $358.1+02.3$ I p& 17 31 58.3 & -29 15 01 & 50.84 & 66.73 & 77.74 & 93.80 & 176.68 \\
PNG $358.4+02.1$ I l& 17 33 40.5 & -29 08 34 & 9.04 & 7.50 & 3.90 & 6.89 & 391.82 \\
PNG $358.7-02.5$ I t& 17 52 36.5 & -31 16 27 & 56.60 & 33.18 & 25.54 & 20.99 & 85.54 \\
PNG $359.2-02.4$ I t& 17 53 39.8 & -30 51 25 & 2.38 & 2.81 & 1.87 & 5.62 & 149.90 \\
PNG $359.4+02.3a$ I t& 17 35 12.0 & -28 09 31 & 4.47 & 4.85 & 6.67 & 10.41 & 187.65 \\
PNG $359.6+04.3$ I l& 17 27 58.4 & -26 53 45 & 3.76 & 4.15 & 4.55 & 6.86 & \nodata \\
PNG $359.7+02.0$ I p& 17 36 56.8 & -28 04 42 & 13.71 & 17.60 & 17.79 & 26.90 & 1156.71 \\
PNG $359.8+03.5$ I p& 17 31 47.8 & -27 09 19 & 8.58 & 8.10 & 9.78 & 20.26 & \nodata \\
\\
{\bf Known PNs}\\
Al 2-B & 17 27 47.06 & -28 11 00.76 & 11.08 & 13.97 & 12.05 & 27.39 & ... \\
Al 2-E & 17 30 14.40 & -27 30 19.41 & 5.19 & 13.05 & 9.96 & 13.98 & ... \\
Al 2-F & 17 30 30.43 & -28 35 54.90 & 0.90 & 1.15 & 0.95 & 2.23 & ... \\
Al 2-G & 17 32 22.67 & -28 14 27.32 & 54.49 & 101.47 & 168.58 & 413.55 & 1251 \\
Al 2-J  & 17 35 35.50 & -27 24 06.50 & 2.43 & 4.45 & 1.58 & 2.69 & 573.0 \\
Al 2-K & 17 36 14.18 & -28 00 46.33 & 10.03 & 12.42 & 13.49 & 22.52 & 419.0 \\
Al 2-O & 17 51 45.29 & -32 03 03.90 & 63.01 & 91.83 & 135.67 & 202.92 & 1301 \\
Al 2-R & 17 53 36.46 & -31 25 25.70 & 5.19 & 6.13 & 4.63 & 8.59 & 672.0 \\
Bl 3-10 & 17 55 20.54 & -29 57 36.13 & 8.11 & 10.15 & 7.78 & 11.16 & 1102 \\
H 1-16  & 17 29 23.39 & -26 26 05.00 & 17.61 & ... & 29.20 & ... & ... \\
H 1-17 & 17 29 40.59 & -28 40 22.12 & 15.44 & 27.34 & 34.14 & 201.90 & ... \\
H 1-18 & 17 29 42.76 & -29 32 50.30 & 14.54 & 19.73 & 36.62 & 158.30 & 4098 \\
H 1-19 & 17 30 02.55 & -27 59 17.54 & 23.83 & 24.21 & 38.19 & 133.18 & ... \\
H 1-20 & 17 30 43.82 & -28 04 06.80 & 9.92 & 17.19 & 15.33 & 83.57 & ... \\
H 1-22 & 17 32 22.14 & -37 57 23.80 & 7.40 & 12.65 & 10.31 & 40.00 & ... \\
H 1-29  & 17 44 13.82 & -34 17 33.05 & 13.21 & 14.62 & 43.78 & 145.56 & 1264 \\
H 1-31 & 17 45 32.10 & -34 33 55.32 & 6.44 & 11.09 & 9.45 & 35.40 & ... \\
H 1-32   & 17 46 06.30 & -34 03 45.40 & 15.60 & 20.59 & 15.41 & 65.04 & 4176 \\
H 1-34  & 17 48 07.57 & -22 46 47.33 & 16.46 & 17.60 & 61.30 & 221.53 & 5535 \\
H 1-40   & 17 55 36.05 & -30 33 32.30 & 18.78 & 30.10 & 48.53 & 489.98 & $>$ 9715 \\
H 1-45  & 17 58 21.87 & -28 14 52.30 & 1409.92 & 2133.31 & 6280.78 & $>$ 3754.65 & 1584 \\
H 1-53  & 18 05 57.43 & -26 29 42.00 & 3.65 & 5.36 & 8.00 & 40.50 & 1089 \\
H 1-6  & 17 06 58.87 & -42 41 09.75 & 22.33 & 28.51 & 29.57 & 72.79 & ... \\
H 1-7   &  17 10 27.39 & -41 52 49.42 & 193.43 & 191.04 & 484.47 & 1380.74 & ...  \\
H 2-10 & 17 27 32.85 & -28 31 06.90 & 5.29 & 8.23 & 5.18 & 12.16 & ... \\
H 2-13 & 17 31 08.08 & -30 10 28.00 & 5.57 & 9.77 & 5.78 & 20.05 & 1331 \\
H 2-20  &  17 45 39.77 & -25 40 00.04 & 45.43 & 50.08 & 43.73 & 93.53 & 3500  \\
H 2-24  & 17 48 36.54 & -24 16 34.80 & 1091.21 & 1512.34 & 3466.65 & $>$ 2831.43 & 3209 \\
H 2-33  &  17 58 12.54 & -31 07 51.10 & 4.23 & 5.57 & 3.36 & 19.49 & ... \\
HDW 8  & 17 31 47.47 & -28 42 03.37 & 46.56 & 60.90 & 49.36 & 150.90 & 2297 \\
Hb 4    &  17 41 52.76 & -24 42 08.07 & 48.36 & 79.03 & 75.62 & 261.68 & ...  \\
Hb 6    &  17 55 07.02 & -21 44 39.98 & 95.71 & 153.35 & 111.41 & 412.57 & 14660  \\
He 2-149  & 16 14 24.27 & -54 47 38.82 & 2.47 & 4.24 & 2.21 & 10.40 & ... \\
He 2-153&  16 17 14.43 & -53 32 08.39 & 15.67 & 17.09 & 18.49 & 35.71 & ...  \\
He 2-157  & 16 22 14.26 & -53 40 54.09 & 8.60 & 12.99 & 10.68 & 45.39 & ... \\
He 2-169&  16 34 13.33 & -49 21 13.20 & 37.56 & 62.84 & 76.82 & 273.96 & 1750  \\
He 2-250& 17 34 54.71 & -26 35 56.92 & 7.05 & 13.68 & 6.56 & 31.51 & ... \\
He 2-262& 17 40 12.84 & -26 44 21.90 & 10.11 & 14.91 & 9.75 & 22.58 & 1405 \\
IC 4673 &  18 03 18.41 & -27 06 22.61 & 16.03 & 46.52 & 17.74 & 137.92 & ...  \\
IRAS 17218-3126  & 17 25 03.47 & -31 28 38.50 & 4.58 & 9.66 & 4.90 & 27.24 & 1072 \\
IRAS 18023-2513 & 18 05 25.51 & -25 13 37.23 & 12.51 & 18.50 & 13.44 & 42.92 & 2445 \\
JaFu 1  & 17 43 57.38 & -26 11 53.98 & 4.46 & 6.41 & 6.25 & 11.72 & 79.0 \\
K 5-10  & 17 41 24.52 & -26 03 53.40 & 29.97 & 34.83 & 35.71 & 60.05 & 776.0 \\
K 5-13  & 17 43 39.44 & -25 36 42.51 & 4.65 & 8.66 & 4.97 & 17.71 & 1335 \\
K 5-16  & 17 45 28.31 & -25 38 10.41 & 3.27 & 4.98 & 3.46 & 12.19 & 139.0 \\
K 5-19  & 17 49 51.28 & -23 27 44.27 & 4.46 & 8.29 & 4.72 & 10.76 & 715.0 \\
K 6-12   & 17 47 17.80 & -33 15 39.00 & 3.37 & 3.99 & 3.45 & 7.30 & 31.0 \\
K 6-14  & 17 48 28.47 & -24 41 25.07 & 8.26 & 9.59 & 7.53 & 21.04 & 1403 \\
KFL 1   & 17 59 15.59 & -30 02 47.15 & 4.05 & 5.94 & 3.22 & 6.98 & ... \\
KFL 2   & 18 00 59.92 & -28 16 19.80 & 3.16 & 2.74 & 1.77 & 2.04 & 182.0 \\
KFL 3    & 18 02 52.93 & -31 23 58.49 & ... & 1.51 & ... & 3.33 & ... \\
KFL 4   & 18 02 51.67 & -27 40 59.60 & 3.85 & 2.91 & 2.39 & 2.05 & 79.0 \\
KFL 5   & 18 03 53.66 & -29 51 21.89 & 31.81 & 56.71 & 85.26 & 225.22 & ... \\
M 1-27  &  17 46 45.45 & -33 08 35.06 & 55.41 & 66.33 & 95.00 & 354.99 & 19390  \\
M 1-31  &  17 52 41.44 & -22 21 57.00 & 29.36 & 40.84 & 64.70 & 252.59 & 9210  \\
M 1-35  & 18 03 39.30 & -26 43 33.90 & 21.14 & 33.09 & 28.46 & 149.55 & 5039 \\
M 1-41  &  18 09 29.90 & -24 12 23.46 & 125.10 & 203.08 & 243.89 & 648.50 & 13670  \\
M 2-21  & 17 58 09.58 & -29 44 20.10 & 21.98 & 29.13 & 57.71 & 177.23 & 1269 \\
M 2-23  & 18 01 42.64 & -28 25 44.20 & 21.89 & 31.26 & 25.92 & 216.84 & 5688 \\
M 2-26  &  18 03 11.41 & -26 58 30.23 & 4.49 & 7.04 & 4.16 & 22.77 & ...  \\
M 2-46  &  18 46 34.61 & -08 28 02.10 & 14.45 & 16.14 & 44.14 & 125.17 & ...  \\
M 3-10 & 17 27 20.19 & -28 27 51.20 & 14.47 & 28.02 & 18.37 & 57.51 & ... \\
M 3-14  & 17 44 20.62 & -34 06 40.60 & 12.62 & 19.91 & 24.61 & 107.50 & 2961 \\
M 3-16 & 17 52 46.05 & -30 49 34.42 & 6.41 & 10.85 & 8.32 & 15.67 & 531.0 \\
M 3-19  & 17 58 19.34 & -30 00 39.32 & 5.36 & 5.71 & 4.75 & 18.12 & 1014 \\
M 3-20  & 17 59 19.35 & -28 13 48.20 & 6.25 & 10.10 & 7.15 & 17.05 & 1581 \\
M 3-22  & 18 02 19.24 & -30 14 25.38 & 1.59 & 4.68 & 3.62 & 6.78 & ... \\
M 3-24  & 18 07 53.91 & -25 24 02.71 & 5.27 & 9.28 & 7.05 & 37.26 & 1154 \\
M 3-46 & 17 55 05.79 & -31 12 16.03 & 7.79 & 8.29 & 7.16 & 18.76 & 140.0 \\
M 3-47  & 17 57 43.37 & -30 02 29.91 & 1.93 & 2.01 & 1.83 & 4.08 & 48.0 \\
M 3-48  &  17 59 56.82 & -31 54 27.46 & 6.58 & 5.35 & 5.54 & 8.23 & ...  \\
M 3-8  & 17 24 52.15 & -28 05 54.61 & 51.30 & 37.21 & 30.33 & 91.70 & ... \\
M 4-10  & 18 34 13.85 & -13 12 24.70 & 9.75 & 16.31 & 12.76 & 41.27 & ... \\
M 4-4   &  17 28 50.29 & -30 07 45.10 & 95.04 & 62.44 & 46.55 & 54.99 & 1120  \\
MaC 1-10 & 18 09 12.88 & -25 04 33.27 & 175.82 & 288.86 & 588.34 & 2076.00 & $>$ 11518 \\
MeWe 1-6  & 16 31 06.65 & -50 26 38.08 & 5.53 & 9.86 & 10.40 & 12.51 & ... \\
Mz 2    &  16 14 32.42 & -54 57 04.20 & 18.24 & 38.94 & 40.98 & 105.89 & ...  \\
NGC 6302&  17 13 44.21 & -37 06 15.94 & 687.77 & 2174.08 & 3400.33 & $>$19020.20 & 252810  \\
NGC 6578&  18 16 16.52 & -20 27 02.67 & 306.55 & 295.80 & 253.76 & 475.52 & ... \\
PN 1824-1410  & 18 27 13.51 & -14 08 34.70 & 25.70 & 43.09 & 33.55 & 31.72 & ... \\
Pe 1-15 &  18 46 24.49 & -07 14 34.57 & 16.75 & 19.77 & 18.52 & 50.80 & ...  \\
Pe 1-6 & 16 23 54.31 & -46 42 15.28 & 6.78 & 10.91 & 4.59 & 28.84 & ... \\
Pe 2-10  & 17 53 37.22 & -21 58 41.80 & 14.52 & 13.03 & 8.90 & 14.17 & 192.0 \\
Pe 2-11 &  17 58 31.27 & -27 37 05.80 & 13.14 & 20.84 & 22.13 & 27.21 & ...  \\
Pe 2-12 & 18 01 10.30 & -27 38 19.88 & 15.77 & 10.74 & 8.15 & 8.36 & 952.0 \\
Sa 3-104& 17 58 25.80 & -29 20 49.00 & 57.38 & 114.46 & 208.25 & 584.52 & 2867 \\
SaWe 2 & 17 27 00.19 & -27 40 35.11 & 8.43 & 8.50 & 11.86 & 26.50 & ... \\
ShWi 1  & 18 02 25.85 & -29 25 05.40 & 0.62 & 0.77 & 0.75 & 2.28 & ... \\
Th 3-10  & 17 24 40.90 & -30 51 59.60 & 14.71 & 26.43 & 26.18 & 80.20 & 2731 \\
Th 3-11 & 17 24 26.33 & -31 43 19.80 & 12.85 & 13.55 & 21.07 & 64.47 & 3148 \\
Th 3-13  & 17 25 19.38 & -29 40 42.00 & 23.17 & 36.26 & 77.98 & 361.50 & $>$ 5871 \\
Th 3-19 & 17 28 41.79 & -28 27 19.32 & 5.43 & 7.32 & 6.24 & 16.86 & ... \\
Th 3-23& 17 30 21.36 & -29 10 12.70 & 13.28 & 27.72 & 22.96 & 33.99 & 1686 \\
Th 3-24  & 17 30 51.35 & -30 17 12.49 & 4.13 & 4.96 & 4.11 & 6.75 & 85.0 \\
Th 3-25  & 17 30 46.81 & -27 05 58.00 & 7.90 & 11.12 & 6.82 & 17.35 & ... \\ 
Th 3-26 & 17 31 09.29 & -28 14 50.43 & 3.75 & 11.05 & 9.14 & 20.10 & $>$ 2495 \\
Th 3-33  & 17 35 48.12 & -27 43 20.38 & 21.28 & 19.27 & 95.83 & 241.35 & 3545 \\
Th 4-3   & 17 48 37.39 & -22 16 48.79 & 4.50 & 5.31 & 5.71 & 17.06 & 1942 \\
Th 4-7   & 17 52 22.57 & -21 51 13.43 & 2.93 & 6.63 & 8.93 & 14.51 & 695.0 \\
Th 4-9   & 17 56 00.60 & -19 29 26.70 & 37.00 & 70.33 & 118.73 & 375.53 & ... \\
Vd 1-5 & 16 51 33.58 & -40 02 56.01 & 1.77 & 3.95 & 2.22 & 4.87 & ... \\
\enddata
\tablenotetext{a}{The `I' and `II' after the PN designations represent the
MASH I and MASH II, respectively. The `t', `l', and `p' represent 
true, likely, and possible PNs, respectively, as assigned in the
MASH I\&II catalogue.}
\tablenotetext{b}{The colon represents uncertain detection. For
 saturating sources, the lower limits of fluxes are given.}
\end{deluxetable}

\begin{deluxetable}{lcc@{\extracolsep{0.1in}}cccc}
\tabletypesize{\scriptsize} \tablecaption{Flux measurements from the AKARI point source catalogue\label{akari}} \tablewidth{0pt}
\tablehead{
Objects & \multicolumn{2}{c}{IRC flux (Jy)} &  \multicolumn{4}{c}{FIS flux (Jy)}\\
\cline{2-3}\cline{4-7}
&
\colhead{9$\micron$} & 
\colhead{18$\micron$} & 
 \colhead{65$\micron$} &
\colhead{90$\micron$} & 
\colhead{140$\micron$} & 
\colhead{160$\micron$}}
\startdata
{\bf MASH PNs}\\
PNG $000.4+04.4$ & \nodata & 0.943   & \nodata & \nodata & \nodata & \nodata \\
PNG $001.5+03.1$ & \nodata & 0.754   & 2.396: & 2.963 & \nodata & \nodata \\
PNG $001.7+03.6$ & \nodata & \nodata & 0.932: & 0.94 & 0.491: & 0.611: \\
PNG $001.7-02.6$ & \nodata & \nodata & 1.173: & 1.77 & \nodata & \nodata \\
PNG $002.3-01.7$ & \nodata & 0.944   & \nodata & \nodata & \nodata & \nodata \\
PNG $002.9-02.7$ & 0.169 & 0.37      & \nodata & \nodata & \nodata & \nodata \\
PNG $003.1-01.6$ & \nodata & \nodata & 4.048: & 6.575 & 4.924: & 6.003: \\
PNG $007.5-02.4$ & \nodata & 0.705   & \nodata & \nodata & \nodata & \nodata \\
PNG $007.8+01.2$ & \nodata & 0.253   & \nodata & \nodata & \nodata & \nodata \\
PNG $009.4-01.2$ & \nodata & 1.388   & \nodata & \nodata & \nodata & \nodata \\
PNG $009.8-01.1$ & \nodata & 0.554   & 4.035: & 4.696 & 11.948 & \nodata \\
PNG $010.2+02.4$ & \nodata & \nodata & 3.122: & 1.417 & 6.395 & 2.206: \\
PNG $010.2+02.7$ & \nodata & 0.844   & \nodata & \nodata & \nodata & \nodata \\ 
PNG $010.6+02.4$ & \nodata & 0.5     & 1.855: & 1.416 & 4.593: & \nodata \\
PNG $010.7-02.3$ & \nodata & 0.55    & 4.957: & 3.122: & 6.939 & 3.708: \\
PNG $011.0+01.4$ & \nodata & \nodata & 0.686: & 3.709 & 8.454 & 9.32: \\
PNG $011.0-02.9$ & \nodata & \nodata & \nodata & 2.609 & \nodata & 5.206: \\
PNG $017.6+02.6$ & \nodata & 1.864   & \nodata & \nodata & \nodata & \nodata \\
PNG $024.2+01.8$ & \nodata & 0.249   & 0.794: & 1.249 & \nodata & \nodata \\
PNG $025.6+02.8$ & \nodata & 0.693   & \nodata & \nodata & \nodata & \nodata \\
PNG $029.0+02.2$ & \nodata & 0.925   & \nodata & \nodata & \nodata & \nodata \\
PNG $030.2+01.5$ & \nodata & 0.204   & \nodata & \nodata & \nodata & \nodata \\
PNG $329.8-03.0$ & \nodata & \nodata & 1.918 & 1.541 & 4.299: & 3.389: \\ 
PNG $330.1+02.6$ & \nodata & 0.863   & \nodata & \nodata & \nodata & \nodata \\
PNG $334.0+02.4$ & \nodata & 0.145   & 3.40: & 1.749 & \nodata & 0.261: \\
PNG $334.4+02.3$ & \nodata & 0.424   & 0.62: & 1.664 & 1.425: & \nodata \\
PNG $335.4-01.9$ & 0.192 & 0.712     & \nodata & 7.93 & 3.24 & 4.735 \\
PNG $335.8-01.6$ & 0.264 & 0.949     & \nodata & \nodata & \nodata & \nodata \\
PNG $335.9-01.3$ & 0.485 & 5.498     & 10.406 & 8.67 & 3.272: & 5.604: \\
PNG $341.9-02.8$ & \nodata & 0.337   & \nodata & \nodata & \nodata & \nodata \\
PNG $344.0+02.5$ & \nodata & 0.426   & \nodata & \nodata & \nodata & \nodata \\
PNG $350.4+02.0$ & \nodata & 1.655   & \nodata & \nodata & \nodata & \nodata \\
PNG $350.8+01.7$ & \nodata & 2.48    & 5.81: & 5.519 & 13.89 & 1.725: \\
PNG $355.0+02.6$ & 0.231 & 3.144     & \nodata & \nodata & \nodata & \nodata \\
PNG $355.2-02.0$ & \nodata & 0.362   & \nodata & \nodata & \nodata & \nodata \\
PNG $356.0+02.8$ & \nodata & 0.584   & \nodata & \nodata & \nodata & \nodata \\
PNG $356.5-01.8$ & \nodata & 0.886   & \nodata & \nodata & \nodata & \nodata \\
PNG $357.3-02.0$ & \nodata & 1.199   & 9.077 & 3.472 & 2.255: & \nodata \\
PNG $357.8+01.6$ & \nodata & 0.631   & \nodata & \nodata & \nodata & \nodata \\
PNG $357.9+01.7$ & \nodata & \nodata & 0.671: & 2.793 & 10.986: & 9.99: \\
PNG $358.0-02.4$ & \nodata & 0.50    & \nodata & \nodata & \nodata & \nodata \\
PNG $358.1+02.3$ & \nodata & 0.173   & \nodata & \nodata & \nodata & \nodata \\
PNG $359.7+02.0$ & \nodata & 1.012   & \nodata & \nodata & \nodata & \nodata \\
PNG $359.8+03.5$ & \nodata & 0.881   & \nodata & \nodata & \nodata & \nodata \\
\\
{\bf Known PNs}\\
Al 2-E & ... & 0.873 & 1.227: & 2.023 & ... & 0.023: \\
Al 2-G & ... & 1.013 & ... & ... & ... & ... \\
Al 2-O & 0.241 & 0.582 & ... & ... & ... & ... \\
Al 2-R & ... & 0.306 & ... & ... & ... & ... \\
Bl 3-10& ... & 0.357 & ... & ... & ... & ... \\
H 1-16 & 0.272 & ... & ... & ... & ... & ... \\
H 1-17 & 0.617 & 6.259 & ... & ... & ... & ... \\
H 1-18 & 0.336 & 2.667 & ... & ... & ... & ... \\
H 1-19 & 0.185 & 2.864 & 5.167 & 4.579 & ... & 1.333: \\
H 1-20 & ... & 1.924 & 3.73 & 3.436 & ... & ... \\
H 1-22 & ... & 1.517 & ... & ... & ... & ... \\
H 1-29 & 0.174 & ... & ... & ... & ... & ... \\
H 1-32  & 0.266 & ... & ... & ... & ... & ... \\
H 1-34 & 0.352 & ... & ... & ... & ... & ... \\
H 1-40 & 1.255 & 10.494 & 8.689 & 7.947 &3.121: & 2.002: \\
H 1-45 & 5.230 & 2.443 & ... & ... & ... & ... \\
H 1-53 & ... & 0.651 & ... & ... & ... & ... \\
H 1-6  & ... & 0.670 & ... & ... & ... & ... \\
H 1-7    & 1.892 & 13.029 & ... & 42.37: & 33.76 & 10.955: \\
H 2-10 & ... & 1.045 & ... & ... & ... & ... \\
H 2-13 & ... & 1.052 & ... & ... & ... & ... \\
H 2-20   & ... & 1.474 & ... & ... & ... & ...  \\
H 2-24 & ... & 3.535 & ... & ... & ... & ... \\
H 2-33   & ... & 0.393 & 1.633: & 2.283 & ... & ... \\
HDW 8  & 0.728 & 2.139 & ... & ... & ... & ... \\
Hb 4     & ... & 5.449 & 13.541 & 14.269 & 0.952: & 1.287: \\
Hb 6     & 1.144 & ... & 20.768: & 14.246 & 6.261: & 2.636: \\
He 2-149 & ... & 0.542 & ... & 1.020 & ... & ... \\
He 2-153 & ... & 0.298 & ... & ... & ... & ... \\
He 2-157 & ... & 1.951 & 2.612: & 2.420 & ... & 0.963: \\
He 2-169 & 0.338 & 1.697 & 12.73 & 11.83 & 14.61 & 18.50: \\
He 2-250 & ... & 0.728 & 2.625 & 3.213 & ... & 0.965: \\
He 2-262 & ... & 1.082 & ... & ... & ... & ... \\
IC 4673  & ... & 3.057 & 9.901: & 9.916 & 6.940: & 6.241: \\
IRAS 17218-3126 & ... & 0.533 & ... & ... & ... & ... \\
IRAS 18023-2513 & 0.156 & 1.668 & 1.296: & 5.988 & 1.724: & ... \\
K 5-10 & ... & 0.693 & ... & ... & .... & ... \\
K 5-13 & ... & 0.658 & ... & ... & ... & ... \\
K 6-14 & ... & 0.627 & ... & ... & ... & ... \\
KFL 5    & 0.468 & 0.800 & ... & ... & ... & ... \\
M 1-27   & 0.446 & ... & ... & ... & ... & ... \\
M 1-31   & 0.435 & 5.727 & ... & ... & ... & ... \\
M 1-35 & 0.514 & 3.572 & ... & ... & ... & ... \\
M 1-41   & 1.320 & 5.912 & ... & ... & ... & ... \\
M 2-21 & 0.305 & 0.789 & ... & ... & ... & ... \\
M 2-23 & ... & 6.155 & ... & ... & ... & ... \\
M 2-26   & ... & 0.398 & ... & ... & ... & ... \\
M 2-46   & 0.111 & 0.455 & 2.976 & 4.938 & 2.008: & 2.894: \\
M 3-10 & 0.311 & 2.611 & 2.650: & 2.658 & ... & 2.107: \\
M 3-14 & 0.344 & 1.889 & ... & ... & ... & ... \\
M 3-19 & ... & 0.800 & ... & ... & ... & ... \\
M 3-20 & 0.091 & ... & 1.786: & 1.353 & ... & ... \\
M 3-22 & ... & 0.451 & 1.334: & 1.319 & 0.941:& 0.998: \\
M 3-24 & ... & 0.887 & 5.998 & 3.891 & 0.071: & ... \\
M 3-48   & ... & ... & 0.104: & 0.419 & ... & ... \\
M 3-8  & 0.297 & ... & ... & ... & ... & ... \\
M 4-10 & ... & 1.920 & ... & ... & ... & ... \\
M 4-4    & 0.088 & ... & ... & ... & ... & ... \\
MaC 1-10 & 2.791 & 12.058 & ... & ... & ... & ... \\
Mz 2     & ... & 1.763 & 6.39 & 7.386: & 2.8: & 0.136: \\
NGC 6302 & ... & 156.07 & 670.18 & 304.59 & 188.33 & 201.675 \\
NGC 6578 & 1.357 & ... & 20.146 & 26.884: & 10.368: & 4.589: \\
Pe 1-15  & 0.079 & 0.722 & 2.534 & 2.029 & ... & ... \\
Pe 1-6 & 0.191 & 0.823 & ... & 4.576 & ... & ... \\
Sa 3-104 & 0.866 & 2.315 & ... & ... & ... & ... \\
SaWe 2 & ... & ... & 0.562 & 2.44: & 1.572 & ... \\
Th 3-10 & 0.152 & 1.635 & ... & ... & ... & ... \\
Th 3-11& ... & 1.542 & ... & ... & ... & ... \\
Th 3-13 & ... & 4.774 & 8.348 & 2.537: & ... & ... \\
Th 3-19& ... & 0.967 & ... & ... & ... & ... \\
Th 3-23& ... & 0.847 & ... & ... & ... & ... \\
Th 3-25& ... & 0.841 & 0.043: & 1.042 & 0.528: & 2.846: \\
Th 3-26& ... & ... & 2.622: & 2.337 & ... & ... \\
Th 3-33& ... & 2.064 & ... & ... & ... & ... \\
Th 4-3 & ... & 1.015 & 1.110: & 1.217 & 0.433: & ... \\
Th 4-7 & ... & 0.276 & ... & ... & ... & ... \\
Th 4-9 & 0.595 & 1.13 & ... & ... & ... & ... \\
Vd 1-5 & ... & 0.139 & ... & ... & ... & ... \\
\enddata

\end{deluxetable}

\clearpage

\pagestyle{empty}

\begin{deluxetable}{lccc@{\extracolsep{0.1in}}ccc@{\extracolsep{0.1in}}
cccc@{\extracolsep{0.1in}}ccccc}
\rotate
\tablecaption{Other flux measurements \label{other}}
\tabletypesize{\tiny}
\tablewidth{9.8in}
\tablehead{
 &
\multicolumn{3}{c}{DENIS} &
\multicolumn{3}{c}{2MASS} &
\multicolumn{4}{c}{{\it MSX}} &
\multicolumn{4}{c}{{\it IRAS}} &
\multicolumn{1}{c}{NVSS}\\
\cline{2-4}\cline{5-7}\cline{8-11}\cline{12-15}
&
\multicolumn{1}{c}{I} &
\multicolumn{1}{c}{J} &
\multicolumn{1}{c}{K} &
\multicolumn{1}{c}{J} &
\multicolumn{1}{c}{H} &
\multicolumn{1}{c}{K} &
\multicolumn{1}{c}{8.28\,$\mu$m} &
\multicolumn{1}{c}{12.13\,$\mu$m} &
\multicolumn{1}{c}{14.65\,$\mu$m} &
\multicolumn{1}{c}{21.3\,$\mu$m} &
\multicolumn{1}{c}{12\,$\mu$m} &
\multicolumn{1}{c}{25\,$\mu$m} &
\multicolumn{1}{c}{60\,$\mu$m} &
\multicolumn{1}{c}{100\,$\mu$m} &
\multicolumn{1}{c}{1.4\,GHz}\\
\multicolumn{1}{c}{Object}&
\multicolumn{1}{c}{(mag)} &
\multicolumn{1}{c}{(mag)} &
\multicolumn{1}{c}{(mag)} &
\multicolumn{1}{c}{(mag)} &
\multicolumn{1}{c}{(mag)} &
\multicolumn{1}{c}{(mag)} &
\multicolumn{1}{c}{(Jy)} &
\multicolumn{1}{c}{(Jy)} &
\multicolumn{1}{c}{(Jy)} &
\multicolumn{1}{c}{(Jy)} &
\multicolumn{1}{c}{(Jy)} &
\multicolumn{1}{c}{(Jy)} &
\multicolumn{1}{c}{(Jy)} &
\multicolumn{1}{c}{(Jy)} &
\multicolumn{1}{c}{(mJy)} 
}
\startdata
{\bf MASH PNs}\\
{  PNG $000.4+04.4$} & {  16.997} & {  14.946} & {  ...} & {  15.166} & {  14.352} & {  13.686} & {  0.070} & {  ...} & {  0.654} & {  1.133} & {  0.383} & {  2.026} & {  3.835} & {  10.490} & {  11.1}\\
{  PNG $001.0+02.2$} & {  15.926} & {  14.312} & {  ...} & {  14.347} & {  13.748} & {  13.336} & {  ...} & {  ...} & {  ...} & {  ...} & {  ...} & {  ...} & {  ...} & {  ...} & {  ...}\\

{  PNG $001.1+02.2$} & {  16.717} & {  ...} & {  ...} & {  15.108} & {  13.583} & {  13.172} & {  ...} & {  ...} & {  ...} & {  ...} & {  ...} & {  ...} & {  ...} & {  ...} & {  ...}\\

{  PNG $001.5-02.8$} & {  ...} & {  ...} & {  ...} & {  14.019} & {  13.064} & {  12.812} & {  ...} & {  ...} & {  ...} & {  ...} & {  ...} & {  ...} & {  ...} & {  ...} & {  ... }\\

{  PNG $001.5+03.1$} & {  ...} & {  ...} & {  ...} & {  14.231} & {  13.155} & {  12.722} & {  ...} & {  ...} & {  ...} & {  ...} & {  $<2.302$} & {  1.825} & {  4.806} & {  $<18.230$} & {  3.5 }\\

{  PNG $002.3-01.7$} & {  15.180} & {  ...} & {  11.971} & {  13.745} & {  13.020} & {  12.089} & {  0.191} & {  ...} & {  ...} & {  1.576} & {  ...} & {  ...} & {  ...} & {  ...} & {  ...}\\

{  PNG $002.5+02.0$} & {  ...} & {  ...} & {  ...} & {  14.549} & {  13.550} & {  13.095} & {  ...} & {  ...} & {  ...} & {  ...} & {  ...} & {  ...} & {  ...} & {  ...} & {  ...}\\

{  PNG $002.9-02.7$} & {  14.010} & {  12.097} & {  10.920} & {  12.234} & {  11.374} & {  11.023} & {  0.139} & {  ...} & {  ...} & {  2.148:} & {  ...} & {  ...} & {  ...} & {  ...} & {  ...}\\

{  PNG $005.8+02.2$a} & {  16.302} & {  14.584} & {  ...} & {  15.203} & {  14.055} & {  13.744} & {  ...} & {  ...} & {  ...} & {  ...} & {  ...} & {  ...} & {  ...} & {  ...} & {  ...}\\

{  PNG $006.0+01.2$} & {  16.869} & {  14.576} & {  ...} & {  14.558} & {  13.633} & {  12.658} & {  ...} & {  ...} & {  ...} & {  ...} & {  ...} & {  ...} & {  ...} & {  ...} & {  ...}\\

{  PNG $006.1+01.5$} & {  14.389} & {  13.384} & {  12.725} & {  13.489} & {  12.948} & {  12.740} & {  0.111} & {  ...} & {  0.978} & {  2.369} & {  $<2.092$} & {  2.910} & {  5.425} & {  46.310} & {  25.2}\\

{  PNG $007.5-02.4$} & {  ...} & {  ...} & {  ...} & {  ...} & {  ...} & {  ...} & {  ...} & {  ...} & {  ...} & {  ...} & {  $<1.391$} & {  1.527} & {  $<17.79$} & {  $<169.50$} & {  9.2}\\

{  PNG $009.4-01.2$} & {  ...} & {  14.395} & {  12.476} & {  14.345} & {  13.129} & {  12.518} & {  ...} & {  ...} & {  ...} & {  ...} & {  ...} & {  ...} & {  ...} & {  ...} & {  18.6}\\

{  PNG $009.8-01.1$} & {  ...} & {  15.358} & {  13.422} & {  14.978} & {  13.894} & {  13.178} & {  ...} & {  ...} & {  ...} & {  ...} & {  ...} & {  ...} & {  ...} & {  ...} & {  3.3}\\

{  PNG $010.0-01.5$} & {  18.170} & {  14.760} & {  13.256} & {  14.890} & {  13.583} & {  13.205} & {  ...} & {  ...} & {  ...} & {  ...} & {  ...} & {  ...} & {  ...} & {  ...} & {  33.4}\\

{  PNG $010.2+02.7$} & {  17.077} & {  15.329} & {  ...} & {  15.356} & {  14.229} & {  13.857} & {  ...} & {  ...} & {  ...} & {  ...} & {  ...} & {  ...} & {  ...} & {  ...} & {  18.3}\\

{  PNG $011.0-02.9$} & {  16.046} & {  14.264} & {  13.058} & {  14.257} & {  13.552} & {  13.247} & {  ...} & {  ...} & {  ...} & {  ...} & {  $<0.362$} & {  0.670} & {  3.112} & {  $<88.61$} & {  15.0}\\

{  PNG $017.6+02.6$} & {  ...} & {  15.578} & {  13.567} & {  16.076} & {  14.196} & {  13.834} & {  0.186} & {  1.155} & {  1.876} & {  2.779} & {  ...} & {  ...} & {  ...} & {  ...} & {  26.6}\\

{  PNG $019.2-01.6$} & {  ...} & {  ...} & {  ...} & {  ...} & {  ...} & {  ...} & {  ...} & {  ...} & {  ...} & {  ...} & {  $<1.521$} & {  1.667} & {  4.330} & {  $<348.10$} & {  15.0}\\

{  PNG $025.6+02.8$} & {  16.864} & {  14.906} & {  13.759} & {  15.036} & {  14.603} & {  13.764} & {  ...} & {  ...} & {  ...} & {  ...} & {  $<0.260$} & {  1.330} & {  $<1.289$} & {  $<17.360$} & {  4.7}\\

{  PNG $029.0+02.2$} & {  16.634} & {  14.393} & {  13.101} & {  14.434} & {  14.091} & {  13.215} & {  ...} & {  ...} & {  ...} & {  ...} & {  $<1.259$} & {  1.611} & {  $<11.460$} & {  $<136.900$} & {  12.2}\\

{  PNG $029.2-01.8$} & {  ...} & {  15.386} & {  ...} & {  15.481} & {  14.294} & {  13.907} & {  ...} & {  ...} & {  ...} & {  ...} & {  ...} & {  ...} & {  ...} & {  ...} & {  ...}\\

{  PNG $329.8-03.0$} & {  16.678} & {  15.359} & {  13.854} & {  15.430} & {  14.756} & {  14.548} & {  ...} & {  ...} & {  ...} & {  ...} & {  ...} & {  ...} & {  ...} & {  ...} & {  ...}\\

{  PNG $330.1+02.6$} & {  16.526} & {  14.374} & {  13.015} & {  14.631} & {  14.195} & {  13.151} & {  ...} & {  ...} & {  ...} & {  ...} & {  $<0.250$} & {  1.994} & {  3.156} & {  $<21.550$} & {  ...}\\

{  PNG $334.0+02.4$} & {  17.184} & {  15.131} & {  ...} & {  15.258} & {  14.549} & {  13.851} & {  ...} & {  ...} & {  ...} & {  ...} & {  ...} & {  ...} & {  ...} & {  ...} & {  ...}\\

{  PNG $334.4+02.3$} & {  17.696} & {  15.548} & {  ...} & {  15.990} & {  14.377} & {  14.564} & {  ...} & {  ...} & {  ...} & {  ...} & {  ...} & {  ...} & {  ...} & {  ...} & {  ...}\\

{  PNG $335.4-01.9$} & {  17.343} & {  14.886} & {  ...} & {  14.351} & {  13.338} & {  12.852} & {  0.169} & {  ...} & {  0.719} & {  ...} & {  $<1.728$} & {  0.979} & {  7.515} & {  $<331.200$} & {  ...}\\

{  PNG $335.8-01.6$} & {  ...} & {  ...} & {  ...} & {  14.309} & {  13.719} & {  13.479} & {  0.218} & {  ...} & {  0.588} & {  ...} & {  0.560} & {  1.730} & {  $<49.04$} & {  $<482.200$} & {  ...}\\

{  PNG $335.9-01.3$} & 15.342 & 13.275 & 11.782 & {  13.275} & {  12.574} & {  11.894} & {  0.364} & {  0.784} & {  1.415} & {  8.175} & {  ...} & {  ...} & {  ...} & {  ...} & {  ...}\\

{  PNG $341.7+02.6$} & {  ...} & {  14.825} & {  13.346} & {  14.864} & {  13.691} & {  13.248} & {  ...} & {  ...} & {  ...} & {  ...} & {  ...} & {  ...} & {  ...} & {  ...} & {  ...}\\

{  PNG $341.9-02.8$} & {  ...} & {  ...} & {  ...} & {  15.907} & {  14.790} & {  14.090} & {  ...} & {  ...} & {  ...} & {  ...} & {  ...} & {  ...} & {  ...} & {  ...} & {  ...}\\

{  PNG $344.0+02.5$} & {  17.194} & {  15.144} & {  ...} & {  15.570} & {  14.878} & {  14.510} & {  ...} & {  ...} & {  ...} & {  ...} & {  ...} & {  ...} & {  ...} & {  ...} & {  ...}\\

{  PNG $344.4+01.8$} & {  ...} & {  ...} & {  ...} & {  15.101} & {  14.422} & {  14.052} & {  ...} & {  ...} & {  ...} & {  ...} & {  ...} & {  ...} & {  ...} & {  ...} & {  ...}\\

{  PNG $345.8+02.4$} & {  ...} & {  ...} & {  ...} & {  16.082} & {  15.622} & {  14.635} & {  ...} & {  ...} & {  ...} & {  ...} & {  ...} & {  ...} & {  ...} & {  ...} & {  ...}\\

{  PNG $345.8+02.7$} & {  16.993} & {  15.674} & {  ...} & {  15.770} & {  15.365} & {  15.392} & {  ...} & {  ...} & {  ...} & {  ...} & {  ...} & {  ...} & {  ...} & {  ...} & {  ...}\\

{  PNG $350.4+02.0$} & {  ...} & {  14.825} & {  13.346} & {  15.038} & {  14.669} & {  14.128} & {  ...} & {  ...} & {  ...} & {  ...} & {  $<1.867$} & {  3.298} & {  6.299} & {  $<154.200$} & {  12.9}\\

{  PNG $350.8+01.7$} & {  ...} & {  14.576} & {  12.408} & {  14.749} & {  13.838} & {  12.593} & {  0.139} & {  0.871} & {  1.221} & {  2.618} & {  $<1.731$} & {  3.715} & {  8.260} & {  $<188.900$} & {  19.6}\\

{  PNG $355.0+02.6$} & {  ...} & {  13.742} & {  11.745} & {  14.230} & {  13.480} & {  12.306} & {  0.217} & {  ...} & {  2.062} & {  4.172} & {  ...} & {  ...} & {  ...} & {  ...} & {  26.1}\\

{  PNG $355.8+01.7$} & {  ...} & {  ...} & {  ...} & {  ...} & {  ...} & {  ...} & {  ...} & {  ...} & {  ...} & {  ...} & {  $<$2.859} & {  1.214} & {  $<5.664$} & {  $<74.050$} & {  14.4}\\

{  PNG $356.0+02.8$} & {  ...} & {  14.466} & {  13.028} & {  14.757} & {  13.634} & {  13.080} & {  ...} & {  ...} & {  ...} & {  ...} & {  ...} & {  ...} & {  ...} & {  ...} & {  ...}\\

{  PNG $356.2+02.5$} & {  ...} & {  13.929} & {  12.156} & {  14.491} & {  13.381} & {  13.440} & {  ...} & {  ...} & {  ...} & {  ...} & {  ...} & {  ...} & {  ...} & {  ...} & {  12.7}\\

{  PNG $356.5-01.8$} & {  16.970} & {  13.421} & {  11.720} & {  13.580} & {  12.246} & {  11.616} & {  ...} & {  ...} & {  ...} & {  ...} & {  $<3.744$} & {  1.921} & {  $<6.375$} & {  $<344.600$} & {  6.1}\\

{  PNG $357.3-02.0$} & {  ...} & {  ...} & {  12.756} & {  14.028} & {  13.008} & {  12.739} & {  ...} & {  ...} & {  ...} & {  ...} & {  ...} & {  ...} & {  ...} & {  ...} & {  4.2}\\

{  PNG $357.5-02.4$} & {  17.322} & {  14.193} & {  12.736} & {  14.231} & {  13.044} & {  12.477} & {  ...} & {  ...} & {  ...} & {  ...} & {  ...} & {  ...} & {  ...} & {  ...} & {  7.3}\\

{  PNG $358.1+02.3$} & {  ...} & {  ...} & {  ...} & {  13.811} & {  12.518} & {  11.313} & {  ...} & {  ...} & {  ...} & {  ...} & {  ...} & {  ...} & {  ...} & {  ...} & {  ...}\\

{  PNG $359.7+02.0$} & {  17.114} & {  13.916} & {  12.302} & {  14.219} & {  12.963} & {  12.448} & {  ...} & {  ...} & {  ...} & {  ...} & {  ...} & {  ...} & {  ...} & {  ...} & {  10.9}\\
\\
{\bf KnownPNs}\\
Al 2-E & ... & ... & ... & ... & ... & ... & ... & ... & ... & ... & $<$ 2.42 & 2.3 & 3.65 & $<$ 15.54 & 14.7\\
Al 2-G & 13.967 & 13.25 & 11.802 & 16.281 & 14.133 & 11.886 & ... & ... & ... & ... & ... & ... & ... & ... & ...\\
Al 2-J & ... & ... & ... & ... & ... & ... & ... & ... & ... & ... & 1.63 & 1.14 & $<$ 1.84 & $<$ 22.03 & 6 \\
Al 2-O & 12.712 & 11.73 & 10.84 & 14.59 & 12.608 & 10.671 & ... & ... & ... & ... & $<$ 2.29 & 1.37 & 3.83 & $<$ 103.9 & 38.3\\
Bl 3-10 & ... & ... & ... & ... & ... & ... & ... & ... & ... & ... & 1.67 & 1.74 & $<$ 15.2 & $<$ 84.85 & 4.7 \\
H 1-16 & 13.142 & 12.547 & 11.638 & 15.367 & 12.893 & 11.662 & 0.218 & ... & 1.971 & 3.74 & 0.682 & 5.95 & 5.13 & $<$ 10.3 & ...\\
H 1-17 & 13.08 & 12.679 & 11.745 & 15.067 & 12.71 & 11.69 & 0.465 & 1.519 & 3.147 & 10.28 & 1.45 & 11.31 & 6.19 & $<$ 33.88 & 6.9\\
H 1-18 & 13.634 & 13.292 & 12.238 & 15.693 & 13.487 & 12.451 & ... & ... & ... & ... & $<$ 2.2 & 5.39 & 6.59 & $<$ 35.69 & 17.6\\
H 1-19 & 12.429 & 11.492 & 11.047 & 14.634 & 12.359 & 10.946 & 0.145 & 1.452 & 0.953 & 4.994 & $<$ 1.97 & 6.17 & 6.5 & $<$ 22.02 & 11.6\\
H 1-20 & 13.704 & 13.502 & 12.445 & ... & ... & ... & ... & ... & ... & ... & $<$ 2.7 & 3.7 & 5.46 & $<$ 20.38 & 26.9 \\
H 1-22 & 13.94 & 13.652 & 12.647 & 15.545 & 13.72 & 12.415 & ... & ... & ... & ... & $<$ 1.45 & 3.35 & 4.96 & $<$ 137.4 & 18.4\\
H 1-29 & 13.542 & 13.104 & 12.339 & 14.901 & 13.481 & 12.425 & ... & ... & ... & ... & ... & ... & ... & ... & ...\\
H 1-31 & 13.274 & 12.88 & 12.825 & ... & ... & ... & ... & ... & ... & ... & $<$ 1.57 & 3.26 & 1.27 & $<$ 11.26 & ...\\
H 1-32 & 12.841 & 12.481 & 11.861 & ... & ... & ... & 0.19 & ... & 1.562 & 4.376 & 0.73 & 5.22 & 1.81 & $<$ 11.24 & ...\\
H 1-34 & 12.805 & 11.975 & 11.625 & ... & ... & ... & 0.309 & ... & 0.921 & 5.346 & $<$ 0.66 & 7.85 & 13.26 & $<$ 16.48 & ... \\
H 1-40 & 12.705 & 12.369 & 11.579 & 14.461 & 12.777 & 11.57 & 1.098 & 2.103 & 4.601 & 16.28 & 2.38 & 18.45 & 11.91 & $<$ 73.48 & 8.5\\
H 1-45 & 10.518 & 8.399 & 6.697 & ... & ... & ... & 4.832 & 5.791 & 3.297 & 2.025 & ... & ... & ... & ... & ... \\
H 1-53 & 14.109 & 13.747 & 12.893 & ... & ... & ... & 0.104 & ... & 1.076 & 2.105 & ... & ... & ... & ... & 7.2 \\
H 1-7 & ... & ... & ... & ... & ... & ... & 1.811 & 2.527 & 8.401 & 17.44 & 3.26 & 25.85 & 58.81 & 33.18 &...\\
H 2-10 & 14.13 & 13.054 & 12.953 & ... & ... & ... & ... & ... & ... & ... & ... & ... & ... & ... & ...\\
H 2-20 & 13.694 & 11.876 & 11.308 & ... & ... & ... & 0.135 & ... & ... & 3.029 & ... & ... & ... & ... & 13.1\\
H 2-24 & 10.3 & 8.276 & 6.944 & 13.634 & 10.171 & 6.929 & 4.156 & 3.943 & 3.993 & 3.103 & ... & ... & ... & ... & 4.5 \\
H 2-33 & ... & ... & ... & ... & ... & ... & ... & ... & ... & ... & 0.76 & 1.1 & 2.81 & $<$59.86 & 8.1\\
HDW 8 & ... & ... & ... & ... & ... & ... & 0.392 & ... & 1.572 & 2.581 & ... & ... & ... & ... & ...\\
Hb 4 & 12.738 & 12.466 & 11.601 & 14.402 & 12.606 & 11.252 & ... & ... & ... & ... & 1.34 & 10.30 & $<$20.85 & 12.78 & 157.6\\
Hb 6 & ... & ... & ... & ... & ... & ... & 1.110 & 0.990: & 7.978 & 15.08 & 1.58 & 22.49 & 27.36 & $<$21.24 & 190.5\\
He 2-149 & 14.648 & 14.543 & 13.849 & 15.551 & ... & 13.475 & ... & ... & ... & ... & $<$ 0.37 & 0.89 & 2.17 & $<$ 114.7 & ...\\
He 2-157 & 13.395 & 13.22 & 12.365 & 14.554 & 13.198 & 12.32 & ... & ... & ... & ... & $<$ 2.81 & 4.1 & 3.2 & $<$ 66.2 & ...\\
He 2-169 & 13.045 & 12.157 & 11.773 & 15.217 & 13.024 & 11.734 & 0.280 & ... & 1.538 & 2.545: & $<$0.87 & 3.18 & 10.02 & $<$390.1 & 53.4\\
He 2-250 & ... & ... & ... & ... & ... & ... & 0.121 & ... & 1.125 & ... & $<$ 2.01 & 2.01 & $<$ 4.51 & 16.15 & 15.3 \\
He 2-262 & 13.779 & 12.632 & 12.55 & 15.581 & 13.647 & 12.619 & ... & ... & ... & ... & 2.19 & 3.0 & $<$ 2.51 & $<$ 27.07 & 24.1 \\
IC 4673 & ... & ... & ... & ... & ... & ... & 0.335 & ... & 1.957 & 4.015: & $<$4.80 & 7.82 & $<$14.11 & $<$44.73 &...\\
IRAS 18023-2513 & ... & ... & ... & ... & ... & ... & ... & ... & ... & ... & $<$ 18.6 & $<$ 11.3 & 8.41 & $<$ 463 & ... \\
K 5-10 & 14.055 & 12.769 & 12.605 & ... & ... & ... & ... & ... & ... & ... & ... & ... & ... & ... & 5.6 \\
K 5-13 & 13.736 & 12.743 & 12.26 & ... & ... & ... & ... & ... & ... & ... & ... & ... & ... & ... & 11.3 \\
K 6-14 & 11.861 & 10.336 & 9.744 & 13.431 & 11.758 & 9.665 & ... & ... & ... & ... & ... & ... & ... & ... & 20.7 \\
KFL 1 & 13.567 & 12.956 & 12.724 & ... & ... & ... & ... & ... & ... & ... & ... & ... & ... & ... & ... \\
KFL 2 & 13.642 & 12.886 & 12.605 & ... & ... & ... & ... & ... & ... & ... & ... & ... & ... & ... & ... \\
KFL 4 & 13.578 & 12.879 & 12.661 & 14.96 & 13.742 & ... & ... & ... & ... & ... & ... & ... & ... & ... & ... \\
KFL 5 & 13.083 & 12.47 & 11.867 & ... & ... & ... & 0.402 & ... & 0.666 & ... & ... & ... & ... & ... & ... \\
M 1-27 & 11.681 & 11.344 & 10.812 & 12.805 & 11.586 & 10.787 & 0.388 & 2.214 & 2.751 & 18.19 & 1.57 & 23.4 & 20.43 & $<$187 & 64.9\\
M 1-31 & 12.751 & 12.509 & 11.495 & 14.342 & 12.455 & 11.507 & 0.419 & 0.792: & 3.095 & 8.135 & 1.17 & 11.68 & 11.07 & $<$34.90 & 28.8\\
M 1-35 & 13.151 & 12.99 & 11.869 & 14.637 & 12.915 & 11.718 & 0.458 & ... & 3.393 & 4.048 & 1.4 & 7.25 & 14.05 & $<$ 154.5 & 54 \\
M 1-41 & 14.181 & 13.999 & 12.369 & ... & ... & ... & ... & ... & ... & ... & 3.27 & 10.74 & 38.18 &  $<$238.6 & ...\\
M 2-21 & 13.2 & 13.079 & 12.016 & ... & ... & ... & 0.232 & 0.711 & 0.834 & ... &  1.02 & 1.51 & $<$ 7.78 & $<$ 99.13 & 20.3 \\
M 2-23 & 12.131 & 12.1 & 11.342 & ... & ... & ... & 0.726 & 1.382 & 3.302 & 8.215 & 1.93 & 9.31 & 0.64 & $<$ 126.7 & ... \\
M 2-46 & 14.378 & 14.218 & 13.304 & 15.183 & 14.152 & 13.027 & ... & ... & ... & ... & $<$0.38 & 0.98 & 4.56 & 10.81 & 13.1\\
M 3-10 & 13.175 & 12.97 & 12.064 & 14.705 & 13.01 & 12.103 & 0.179 & ... & 1.402 & 3.346 & $<$ 0.6 & 5.11 & 4.27 & $<$ 16.23 & 35.2\\
M 3-14 & 13.831 & 13.153 & 12.407 & 15.324 & 13.315 & 12.555 & 0.23 & 1.305 & 1.68 & 2.502 & $<$ 4.43 & 3.31 & 7.33 & $<$ 13.69 & 22.6\\
M 3-16 & ... & ... & ... & ... & ... & ... & ... & ... & ... & ... & $<$ 2.6 & 0.88 & 2.42 & $<$ 45.58 & ...\\
M 3-19 & 12.154 & 11.176 & 10.921 & ... & ... & ... & ... & ... & ... & ... & $<$ 4.44 & 1.47 & 5.09 & $<$ 63.03 & 7.9 \\
M 3-20 & 13.674 & 13.512 & 12.756 & ... & ... & ... & ... & ... & ... & ... & ... & ... & ... & ... & 16.2 \\
M 3-22 & ... & ... & ... & ... & ... & ... & ... & ... & ... & ... & $<$ 5.37 & 2.34 & 1.81 & $<$ 69.59 & 10.1 \\
M 3-47 & 14.043 & 13.356 & 13.043 & ... & ... & ... & ... & ... & ... & ... & ... & ... & ... & ... & ...\\
M 3-8 & 11.232 & 10.059 & 9.687 & 13.397 & 11.12 & 9.631 & ... & ... & ... & ... & 0.67 & 4.05 & 5.01 & $<$ 14.62 & 18 \\
M 4-10 & 13.351 & 13.105 & 12.191 & 15.124 & 13.236 & 12.077 & ... & ... & ... & ... & ... & ... & ... & ... & 12.1\\
MaC 1-10 & ... & ... & ... & ... & ... & ... & ... & ... & ... & ... & 5.92 & 20.62 & 22.64 & $<$ 164.6 & ... \\
Mz 2 & 13.693 & 13.195 & 13.034 & 14.754 & 13.732 & 13.056 & 0.174 & ... & 0.9818 & 1.219: & 0.49 & 4.75 & 8.2 & 11.42 &...\\
NGC 6302 & 11.255 & 10.706 & 9.442 & 12.557 & 10.966 & 9.408 & ... & ... & ... & ... & 32.08 & 335.9 & 849.7 & 537.4 & 1908\\
NGC 6578 & 12.649 & 9.825 & 8.934 & ... & ... & ... & 1.159 & 1.242: & 6.541 & 10.71 & $<$12.8 & 16.1 & 37.31 & $<$355 & 158.2\\
Pe 1-6 & ... & ... & ... & ... & ... & ... & ... & ... & ... & ... & $<$ 4.46 & 1.53 & 5.15 & $<$ 40.91 & ...\\
Pe 2-10 & 12.14 & 11.39 & 10.865 & 13.91 & 12.047 & 10.864 & ... & ... & ... & ... & ... & ... & ... & ... & ...\\
Pe 2-12 & 11.627 & 10.947 & 10.772 & 12.981 & 11.682 & 10.816 & ... & ... & ... & ... & $<$ 2.21 & 1.29 & 6.59 & $<$ 186.3 & ... \\
Sa 3-104 & 13.332 & 12.613 & 11.809 & ... & ... & ... & 0.828 & 1.41 & 1.23 & 2.685 & 1.39 & 3.45 & 1.74 & $<$ 92.57 & ... \\
ShWi 1 & 13.407 & 12.831 & 12.56 & 14.775 & 13.428 & 12.613 & ... & ... & ... & ... & ... & ... & ... & ... & ... \\
Th 3-10 & 14.404 & 13.581 & 12.586 & 17.645 & 14.241 & 12.491 & ... & ... & ... & ... & $<$ 2.62 & 3.16 & 4.81 & $<$ 48.22 & 21.6\\
Th 3-11 & 14.395 & 11.123 & 10.61 & ... & ... & ... & ... & ... & ... & ... & ... & ... & ... & ... & 6.9\\
Th 3-13 & 13.674 & 13.067 & 11.947 & 16.138 & 13.831 & 11.932 & 0.556 & 1.641 & 2.0 & 6.828 & 1.3 & 8.69 & 3.15 & $<$ 4.17 & ...\\
Th 3-19 & 14.147 & 13.527 & 12.87 & ... & ... & ... & ... & ... & ... & ... & 4.51 & 7.12 & $<$ 1.84 & $<$ 20.16 & ...\\
Th 3-23 & 14.059 & 12.821 & 12.38 & 16.634 & 13.863 & ... & ... & ... & ... & ... & ... & ... & ... & ... & 46.2 \\
Th 3-25 & 13.866 & 13.17 & 12.429 & ... & ... & ... & ... & ... & ... & ... & ... & ... & ... & ... & ...\\
Th 3-26 & ... & ... & ... & ... & ... & ... & ... & ... & ... & ... & 2.43 & 4.31 & $<$ 2.63 & $<$ 29.7 & 10.4\\
Th 3-33 & 13.51 & 12.34 & 11.9 & ... & ... & ... & 0.351 & 0.879 & 0.491 & 3.336 & 0.89 & 4.76 & 8.67 & $<$ 30.45 & ...\\
Th 4-3 & 13.583 & 13.223 & 12.782 & 14.925 & 13.702 & 12.862 & ... & ... & ... & ... & $<$ 0.785 & 1.928 & $<$ 2.78 & $<$ 8.731 & ... \\
Th 4-9 & 14.431 & 13.529 & 12.214 & 16.268 & 14.617 & 12.084 & 0.573 & 0.892 & 0.909 & 1.856 & 1 & 1.85 & $<$ 1.41 & $<$ 15.78 & ...\\
Vd 1-5 & 15.548 & 15.103 & 14.569 & 16.985 & 15.262 & ... & ... & ... & ... & ... & ... & ... & ... & ... & ...\\
\enddata
\end{deluxetable}

\end{document}